\documentclass[preprint,12pt]{elsarticle}

\biboptions{sort&compress}

\usepackage{amssymb}
\usepackage{bm}
\usepackage{amsmath}

\usepackage{xcolor}

\usepackage{caption}
\usepackage{subcaption}
\usepackage{booktabs}
\usepackage{media9}

\journal{Combustion and Flame}

\begin{document}

\begin{frontmatter}



\title{FGM Modeling of Thermo-Diffusive Unstable Lean Premixed Hydrogen-Air Flames}


\author[inst1]{Stijn N.J. Schepers}

\affiliation[inst1]{organization={Mechanical Engineering, Eindhoven University of Technology},
            addressline={Groene Loper 3}, 
            city={Eindhoven},
            postcode={5612 AE}, 
            country={the Netherlands}}

\author[inst1]{Jeroen A. van Oijen}


\begin{abstract}

Ultra-lean premixed hydrogen combustion is a possible solution to decarbonize industry, while limiting flame temperatures and thus nitrous oxide emissions. These lean hydrogen/air flames experience strong preferential diffusion effects, which result in thermo-diffusive (TD) instabilities. To efficiently and accurately model lean premixed hydrogen flames, it is crucial to incorporate these preferential diffusion effects into flamelet tabulated chemistry frameworks, such as the Flamelet-Generated Manifold (FGM) method. This is challenging because the preferential diffusion terms in the control variable transport equations contain diffusion fluxes of all species in the mechanism. In this work, a new implementation is presented; the full term is reduced by only considering the most contributing species. When carefully selecting this set of major species, preferential diffusion fluxes along the flame front, i.e., cross-diffusion, can be captured. This is particularly important for manifolds that include heat loss effects, where enthalpy is one of the control variables. The diffusion of the H-radical has a significant contribution to the enthalpy transport equation, and cross-diffusion of the H-radical is non-negligible. Two manifolds, without and with heat loss effects, and the set of major species are analyzed in an \textit{a-priori} and \textit{a-posteriori} manner. Simulations of TD unstable hydrogen-air flames with detailed chemistry and several FGM models show that accurately capturing cross-diffusion of enthalpy is important for correctly predicting the flame shape and dynamics.

\end{abstract}

\begin{graphicalabstract}
\includegraphics[width=\textwidth]{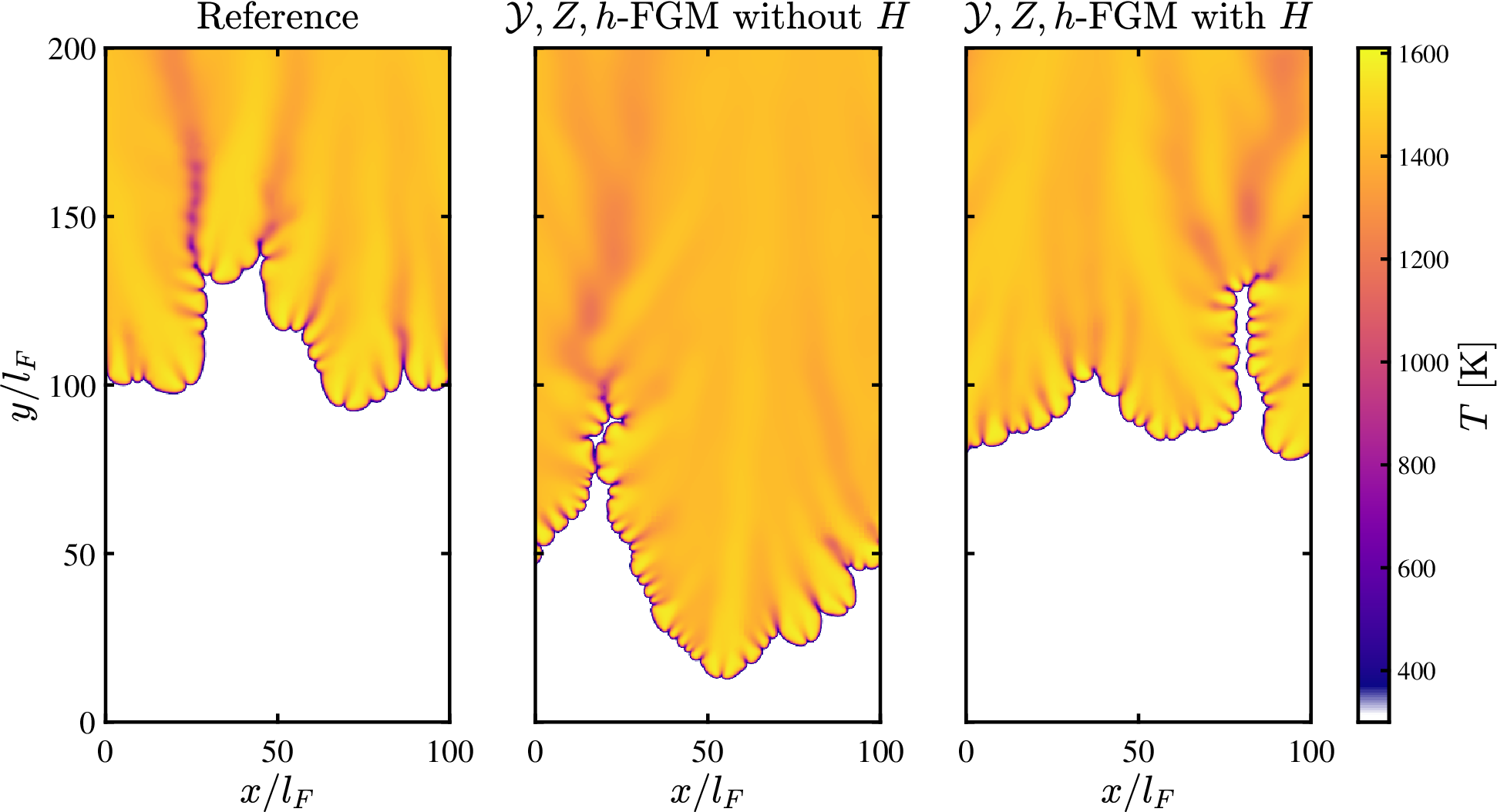}
\end{graphicalabstract}

\begin{highlights}
\item A thermo-diffusive unstable flame is simulated with FGM tabulated chemistry
\item A simplified preferential diffusion approach is presented which is able to accurately predict the typical flame fingers using different manifolds
\item An analysis of species contributing most to preferential diffusion has shown that the H-radical plays a critical role in cross-diffusion of enthalpy, i.e, when heat loss is included in the manifold.
\end{highlights}


\end{frontmatter}


\section{Introduction}
\label{sec:introduction}

Deploying hydrogen as an energy carrier is a key strategy for achieving a sustainable, net-zero CO$_2$ emissions future. In industries where high temperatures are required, e.g., energy generation with gas turbines, hydrogen can be used as an alternative to hydro-carbon based fuels. Due to hydrogen's wide flammability limit, lean combustion mixtures can be used, which is beneficial for reducing the firing temperature and thus limiting NO$_x$ emissions. Lean premixed hydrogen-air flames have intrinsic unstable flame front dynamics, which affect flame propagation \cite{Altantzis2011DetailedFlames,Altantzis2012HydrodynamicFlames,Kadowaki2005TheInstability,Berger2019CharacteristicFlames,Berger2023FlameFlames,Creta2020PropagationInstabilities,Kadowaki2005NumericalInteraction,Wen2024ThermodiffusivelyPatterns}.

The two major mechanisms behind these instabilities are the Darrieus-Landau (DL) instability, which is caused by the density jump over the flame front, and the Thermo-Diffusive (TD) instability due to the imbalance between heat flux leaving the reaction zone and mass flux entering the reaction zone. The coupling between these two effects has been studied by Matalon et al.~\cite{Matalon1982FlamesDiscontinuities,Matalon2003HydrodynamicOrders} where they derived an asymptotic theory, and found that the TD effects only have a destabilizing influence when the effective Lewis number $Le_{\mathrm{eff}}$ is smaller than a critical value $Le_{\mathrm{eff}}^*$. The effective Lewis number depends on the deficient reactant, which is the fuel in the lean combustion regime. Hydrogen has a Lewis number which is significantly lower than unity, which results in enhanced flame front instabilities by TD effects in lean mixtures.

Lean laminar TD unstable hydrogen-air flames have been investigated by numerous sources in the literature \cite{Altantzis2011DetailedFlames,Altantzis2012HydrodynamicFlames,Kadowaki2005TheInstability,Berger2019CharacteristicFlames,Berger2023FlameFlames,Creta2020PropagationInstabilities,Kadowaki2005NumericalInteraction,Wen2024ThermodiffusivelyPatterns}. Altantzis et al.~\cite{Altantzis2011DetailedFlames,Altantzis2012HydrodynamicFlames} studied the initial linear growth of superimposed perturbations and the long-term nonlinear evolution of instabilities by 2-D numerical simulations with single-step chemistry and detailed transport for varying domain sizes. They show that positively-curved segments of the flame experience increased reactivity, while the reactivity in negatively-curved segments decreases. Furthermore, that the domain size significantly influences the stability and structure of the flame. Kadowaki et al.~\cite{Kadowaki2005TheInstability} performed 2-D numerical simulations of a TD unstable hydrogen-air flame and also found that the size of the computational significantly influences the flame dynamics. Berger et al.~\cite{Berger2019CharacteristicFlames} increased the domain size even further in attempt to rule out these confinement effects on the flame front dynamics. They found that for a sufficiently large domain size, the flame consumption speed becomes independent of the domain size. In later work, Berger et al.~\cite{Berger2023FlameFlames} quantified the contribution of each instability mechanism (DL and TD) which confirms the destabilizing effect of the TD instabilities for lean hydrogen-air mixtures, in line with the theory of Matalon et al.~\cite{Matalon2003HydrodynamicOrders}. Creta et al.~\cite{Creta2020PropagationInstabilities} showed that hydrodynamic (DL) instabilities lead to large-scale cusp-like flame structures, while TD instabilities introduce small-scale corrugations. The interaction of these instabilities results in complex flame morphologies and enhanced propagation speeds. In later work of Kadowaki et al.~\cite{Kadowaki2005NumericalInteraction} they performed 2- and 3-D numerical simulations of TD unstable cellular hydrogen-air flames and found that in 3-D the increment of flame surface area and flame velocity is about twice that of 2-D cellular flames, which suggests that the TD effects are even more prominent in 3-D. This has also recently been confirmed by Wen et al.~\cite{Wen2024ThermodiffusivelyPatterns}, who did a large-scale 3-D numerical simulation of a laminar lean premixed hydrogen-air flame. They quantified the differences between 2- and 3-D TD unstable flames for sufficiently large domain sizes to rule out the influence of domain size, and found that for 3-D flames the reactivity is enhanced due to higher peak curvature values. Howarth et al.~\cite{Howarth2022AnFlames,Howarth2023Thermodiffusively-unstablePoints} performed 2- and 3-D simulations of lean TD unstable hydrogen flames and identified a single empirical instability parameter $\omega_2$ that relates the multi-dimensional unstable flame speed and thickness to their laminar one-dimensional values. Aspden et al.~\cite{Aspden2016,Aspden2017} investigated diffusive effects in turbulent lean premixed methane and hydrogen flames, and found that the H-radical plays a particularly prominent role in hydrogen flames. The rapid diffusion of atomic hydrogen has a strong impact on flame structure; their three-dimensional DNS results show that neglecting this effect leads to thinner flames, higher peak reaction rates, and a significantly altered coupling between flame curvature and local reactivity.

Most of these studies have been performed with either one-step chemistry, which has limited accuracy, or detailed chemistry which is computationally much more demanding. This limits these models to mostly fundamental research and academic test cases. Flamelet-based reduced chemistry models, such as Flamelet-Generated Manifolds (FGM) \cite{vanOijen2000ModellingManifolds,VanOijen2016}, Flamelet/Progress Variable (FPV) \cite{Pierce2004Progress-variableCombustion} or the Flame Prolongation for ILDM (FPI) model \cite{Gicquel2000LAMINARDIFFUSION}, have proven to give accurate and computationally efficient results, which enables combustion modeling for industrial cases. Despite the small chemical mechanisms associated with pure hydrogen combustion, reduced chemistry models generally still offer significant speedup. Furthermore, in practical applications, a part methane or ammonia is often blended with hydrogen, necessitating the use of large chemical mechanisms due to the significant increase in intermediate species. Deploying reduced chemistry models in such cases generally result in speedups of several orders of magnitude. Therefore, the development of accurate reduced chemistry models for hydrogen combustion remains an important field of research.

The main idea behind flamelet-based reduced chemistry methods, is to parameterize the complex combustion chemistry into a reduced set of transported control variables. All chemical and thermo-physical parameters are pre-computed by one-dimensional flamelet simulations and stored in a low-dimensional manifold. During runtime, transport equations for the control variables are solved and these field values are used to retrieve the chemical and thermo-physical parameters from the manifold. The set of control variables should be carefully selected to well represent all relevant physical phenomena. A progress variable is selected, which is the control variable that governs the global chemical reaction progress. To account for heat-losses in the system, enthalpy is often selected as an additional control variable. Finally, to account for local changes in mixture composition caused by the non-unity Lewis number effects associated with lean premixed hydrogen combustion, a mixture fraction is added as a control variable, as introduced by van Oijen and de Goey \cite{vanOijen2002ModellingMethod}. As stated before, TD instabilities are a result of these non-unity Lewis number effects. These effects are also referred to as preferential diffusion, and numerous attempts to incorporate these effects efficiently in reduced chemistry models have been made in the literature \cite{DeSwart2010InclusionManifolds,Donini2015DifferentialFlames,Regele2013AFlames,Schlup2019ReproducingFlames,Abtahizadeh2015DevelopmentFlames,Kai2023LESEffect,Zhang2021LargeStretch,Zhang2023Large-eddyManifolds,Mukundakumar2021AFlames,Bottler2022FlameletMixtures,Bottler2023FlameletFlames}. Some of these are based on the assumption that the preferential diffusion flux of the control variables only occurs in the gradient direction of one or more control variables \cite{DeSwart2010InclusionManifolds,Donini2015DifferentialFlames,Regele2013AFlames,Schlup2019ReproducingFlames}. De Swart et al.~\cite{DeSwart2010InclusionManifolds} proposed a model where it is assumed that locally the control variables are a function of solely progress variable, which reduces the full preferential diffusion term to a pre-computed preferential diffusion coefficient multiplied by the progress variable gradient. This approach is memory efficient, because this eliminated storing all species and their diffusion coefficients in the manifold. Donini et al.~\cite{Donini2015DifferentialFlames} extended this idea to a manifold with three control variables: progress variable, enthalpy and mixture fraction based on the element mass fractions according to Bilger \cite{Bilger1990OnFlames}. This approach only holds when preferential diffusion of the control variables occurs solely in flamelet direction, i.e., when the preferential diffusion flux tangentially along the flame front is (close to) zero. For lean hydrogen-air flames this is not the case, and cross-diffusion of the control variables can not be neglected \cite{Mukundakumar2021AFlames}. Furthermore, when pre-computing the preferential diffusion coefficient, dividing by the control variable gradient, which reduces to zero at its limits, can lead to nonphysical numerical peaks in the coefficients, requiring careful treatment. Regele et al.~\cite{Regele2013AFlames} derived a two-equation model based on a similar assumption as De Swart et al.~\cite{DeSwart2010InclusionManifolds} and Donini et al.~\cite{Donini2015DifferentialFlames}, and Schlup et al.~\cite{Schlup2019ReproducingFlames} extended this to also include the mixture fraction gradient and thermal (Soret) diffusion in the preferential diffusion flux of the control variables. Abtahizadeh et al.~\cite{Abtahizadeh2015DevelopmentFlames} used this approach to predict autoignition of CH$_4$/H$_2$ flames and Kai et al.~\cite{Kai2023LESEffect} and Zhang et al.~\cite{Zhang2021LargeStretch,Zhang2023Large-eddyManifolds} applied these ideas to turbulent flames in a large-eddy simulation (LES) context. Furthermore, Fortes et al.~\cite{Fortes2024AnalysisModel} applied this approach to study TD instabilities for lean premixed hydrogen flames. They study the effect of preheating and higher pressure, as well as the effect of the grid size on the solution using a 2-dimensional manifold with progress variable and mixture fraction as control variables.
In more recent work Mukundakumar et al.~\cite{Mukundakumar2021AFlames} proposed a method to include cross-diffusion by relaxing the assumption of Donini et al.~\cite{Donini2015DifferentialFlames}. They rewrote the preferential diffusion flux of a control variable by moving the summation over the species fluxes inside the gradient operator, which allows to pre-compute and store this term directly in the manifold. This works for constant Lewis numbers, but for a mixture-averaged diffusion model, i.e., where the gradient of the Lewis number is not equal to zero, this still leads to storing the non-constant Lewis number fields of all the species into the manifold. Evaluating the full preferential diffusion term of the control variables including cross-diffusion, mixture averaged coefficients and thermal (Soret) diffusion, necessitates storing all species including their diffusion coefficients in the manifold. This requires a significant amount of memory, especially for larger mechanisms. One could choose to pre-compute the full preferential diffusion term of each control variable, store this one term in the manifold, and incorporate it in the transport equation as a source term. However, when doing this no curvature effects can be taken into account, which are critical when modelling TD instabilities. A different approach has been proposed by Böttler et al.~\cite{Bottler2022FlameletMixtures,Bottler2023FlameletFlames}, deviating slightly from the classical approach. They solve the transport of a selection of major species, H$_2$, O$_2$ and H$_2$O, and enthalpy, then construct a progress variable and mixture fraction from these major species during the CFD simulation to look up in the manifold. It is assumed that the reaction progress and preferential diffusion effects can be captured by these major species only. The three major species diffusion fluxes are used directly in the preferential diffusion term of enthalpy together with a pre-computed closure source term, which is the combined effect of all other species. Böttler et al.~applied this model to a TD unstable, spherically expanding hydrogen-air flame and proposed a new approach with curvature as control variable instead of enthalpy \cite{Bottler2023FlameletFlames}. Incorporating curvature in the manifold seems to enhance accuracy slightly, but discrepancies between detailed chemistry and the flamelet models remain to exist. 

As mentioned earlier, Fortes et al.~\cite{Fortes2024AnalysisModel} used a two-dimensional manifold with a progress variable and mixture fraction as control variables to simulate a lean premixed TD unstable hydrogen flame. However, in most engineering cases, heat losses play an important role, making it important to accurately capture the TD unstable preferential diffusion effects when using a three-dimensional manifold with enthalpy as an additional control variable. Böttler et al.~\cite{Bottler2023FlameletFlames} made an attempt, but his approach still showed significant discrepancies when compared to the detailed chemistry benchmark. They proposed an improved model, which replaces the enthalpy control variable by a curvature control variable, excluding the effect of heat losses again. To the best of the authors' knowledge, no one has successfully simulated a planar TD unstable flame with tabulated chemistry using a three-dimensional manifold that includes heat losses which accurately captures the preferential diffusion effects. 

In this work a preferential diffusion approach is presented which is based upon solving transport equations for the control variables and assuming that the preferential diffusion of all control variables is governed by only a few major species. The selection of major species is analysed for different manifolds and it turns out that the major species presented by Böttler et al.~\cite{Bottler2022FlameletMixtures} might not be the most representative selection. The aim of this paper is to highlight the importance and sensitivity of the modelling choices regarding preferential diffusion of enthalpy by showing that including or excluding a single species to the major species selection can significantly affect the accuracy of the solution.

The paper is structured as follows: in Section \ref{sec:numerical_model} the numerical model is presented, starting with a general description of reactive flows and the selection of the transport models. Subsequently, a transport equation for a general control variable is derived, following with an analysis of the contribution of each chemical species to the preferential diffusion flux of the control variables. Based on this analysis, several FGM models with different selections of major species are created. In Section \ref{sec:results_and_verification} the results are presented, starting with an a-priori analysis of the manifolds, followed by a linear stability analysis of the different models. Subsequently, the models are applied to a laminar planar TD unstable lean hydrogen-air flame similar to the case of Berger et al.~\cite{Berger2019CharacteristicFlames,Berger2023FlameFlames}. The flame dynamics as well as the relevant field variables are analysed for the different models. The emphasis is on the difference between two models based on the same three-dimensional manifolds including heat loss but having a different set of major species. The impact of the major species selection on flame dynamics and their capability to capture the cross-diffusion of enthalpy is evaluated. Finally, in Section \ref{sec:conclusions} the main findings of this work are listed. 



\subsection*{Novelty and Significance Statement}

This work presents a novel preferential diffusion formulation within the FGM framework. This formulation reduces the full term, which includes the diffusion fluxes of all species, to a reduced term with only the most contributing species. This allows for accurately modeling cross-diffusion of the enthalpy control variable for a manifold including heat losses, which is demonstrated to be important to predict flame shape and dynamics. To the best of the authors' knowledge, this is the first successful simulation of a thermo-diffusively unstable flame using a manifold that incorporates heat losses. This enables accurate modeling of flame instabilities in practical applications, where heat losses generally play a significant role. 

\section{Numerical Model}
\label{sec:numerical_model}
In this section, the general mathematical framework for reacting flows, which serves as a base for the detailed chemistry solver, will be derived. Next, the control variable equations are derived followed by an analysis of the contribution of several chemical species on the preferential diffusion on the FGM control variables. Finally, the FGM models that are considered in this work are presented.


\subsection{Reactive Flows}
\label{sec:reactiveflows}
Reacting flows can be mathematically described using conservation equations for mass, momentum, chemical species and energy. In this work, the main focus is on chemistry modelling, so solely the mass conservation of chemical species and energy will be elaborated in this section. The conservation of species, i.e., the species mass fraction $Y_i$, can be written as
\begin{equation}
    \label{eq:speciesconservation}
    \frac{\partial\left(\rho Y_i\right)}{\partial t} + \nabla\cdot\left(\rho \mathbf{u} Y_i\right) = -\nabla\cdot\left(\rho\mathbf{U}_iY_i\right) + \dot{\omega}_i\;, \qquad i\in[1,N_s] \;,
\end{equation}
where $\rho$ represents the fluid density, $\mathbf{u}$ the flow velocity vector, and $\dot{\omega}_i$ the chemical source term of species $i$. Using the Hirschfelder and Curtiss approximation, the diffusion velocity vector of species $i$, denoted with $\mathbf{U}_i$ can be written as


\begin{equation}
    \label{eq:hirschfelder}
    \mathbf{U}_i = -\frac{D_i}{X_i}\nabla X_i -\frac{D_i^T}{T}\nabla T\;, \qquad i\in[1,N_s] \;.
\end{equation}
Here represents $X_i$ the molar concentration, $D_i$ and $D_i^T$ the mixture averaged molecular and thermal diffusion coefficients, respectively, and $T$ the fluid temperature. The first term on the right hand side represents molecular diffusion due to a concentration gradient and the second term thermal diffusion, or commonly referred to as the Soret effect. Assuming that the spatial variation of the mean molar mass is negligible, the molar concentration in Eq. (\ref{eq:hirschfelder}) is directly replaced by the species mass fraction. In this work, the fluid's energy is quantified using total enthalpy $h$. When neglecting the effect of pressure variations, energy production due to viscous dissipation and external energy sources, the conservation of total enthalpy under the low mach assumption can be written as
\begin{equation}
    \frac{\partial (\rho h)}{\partial t} + \nabla\cdot\left(\rho\mathbf{u}h\right) = - \nabla\cdot\bm{q}\;.
\end{equation}
Here represents $\bm{q}$ the heat flux vector, which can be written as

\begin{equation}
\label{eq:fourrier}
    \bm{q} = -\lambda\nabla T + \sum_{i=1}^{N_s}h_i\rho \mathbf{U}_iY_i\;,
\end{equation}
where $\lambda$ and $h_i$ denote the fluid's thermal conductivity and the specific enthalpy of species $i$, respectively. The enthalpy gradient can be written as,
\begin{equation}
\label{eq:enthalpygradient}
\nabla h = c_p\nabla T + \sum_{i=1}^{N_s}h_i\nabla Y_i\;,
\end{equation}
where $c_p$ represents the fluid's thermal heat capacity. Substituting Eq. (\ref{eq:enthalpygradient}) in Eq. (\ref{eq:fourrier}) gives the following definition of the heat flux,
\begin{equation}
\label{eq:prefdifenthalpy}
    \bm{q} = -\frac{\lambda}{c_p}\nabla h + \sum_{i=1}^{N_s}h_i\left(\rho \mathbf{U}_iY_i + \frac{\lambda}{c_p}\nabla Y_i\right)\;.
\end{equation}
The first term on the right hand side represents the Fourier heat diffusion and the second term the enthalpy variations due to preferential diffusion of species. The complete set of equations is closed using the caloric and thermal equations of state, and mixture averaged values for the viscosity $\mu$ and thermal conductivity are used. Furthermore, the abundant species N$_2$ is not computed by solving the conservation equation but via the constraint $\sum_{i=1}^{N_s} Y_i=1$. Finally, a correction velocity is added to the flow velocity in the species conservation equation to ensure mass conservation \cite{ThierryPoinsot2005TheoreticalCombustion}. OpenFOAM's default detailed chemistry solver, \textit{reactingFoam}, is extended to mixture averaged diffusion coefficients, viscosity and conductivity, as well as thermal (Soret) diffusion and the correction velocity term. 

\subsection{FGM Control Variables}
\label{sec:FGMControlVariables}
According to the FGM assumption, a multidimensional flame front can be constructed from a continuous set of one-dimensional flamelets \cite{vanOijen2000ModellingManifolds,VanOijen2016}. The thermo-chemical properties of the fluid are retrieved from a lookup table, and it is assumed that they can be described by the spatial and/or temporal evolution of one or more control variables in the multidimensional simulation. The general transport equation of a control variable $\phi_k$ can be written as
\begin{equation}
    \label{eq:controlvariableconservation}
    \frac{\partial\left(\rho \phi_k\right)}{\partial t} + \nabla\cdot\left(\rho \mathbf{u} \phi_k\right) - \nabla\cdot\left(\frac{\lambda}{c_p}\nabla \phi_k\right)  = -\nabla\cdot \mathbf{J}_{\phi_k} + \dot{\omega}_{\phi_k}\;, \qquad k\in[1,N_{c}] \;,
\end{equation}
where $N_c$ represents the total number of control variables, $\dot{\omega}_{\phi_k}$ the control variable source term and $\mathbf{J}_{\phi_k}$ the control variable preferential diffusion flux, which can be written as
\begin{equation}
    \label{eq:controlvariablediffusionflux1}
    \mathbf{J}_{\phi_k} =  -  \sum_{i=1}^{N_s} c_{\phi_{k,i}}\left(\rho \mathbf{U}_i Y_i + \frac{\lambda}{c_p}\nabla Y_i\right)\;, \qquad k\in[1,N_{c}]\;.
\end{equation}
This expression is similar to Eq. (\ref{eq:prefdifenthalpy}) but made more generic by replacing the specific enthalpy $h_i$ with the specific control variable coefficient $c_{\phi_k,i}$. Substituting the definition of the diffusion velocity, given in Eq. (\ref{eq:hirschfelder}), gives,

\begin{equation}
    \label{eq:controlvariablediffusionflux}
    \mathbf{J}_{\phi_k} = -  \sum_{i=1}^{N_s} c_{\phi_{k,i}}\left[\left(\rho D_i - \frac{\lambda}{c_p}\right)\nabla Y_i + \rho D_i^TY_i\frac{\nabla T}{T} \right]\;, \qquad k\in[1,N_{c}]\;.
\end{equation}
Note that the diffusion flux consists of molecular diffusion and thermal (Soret) diffusion.

The first control variable $\phi_1$ considered is the progress variable $\mathcal{Y}$. A progress variable is continuously increasing from an unburnt $\mathcal{Y}_u$ to a burnt $\mathcal{Y}_b$ state to describe the global chemical reaction progress. It is often constructed as a linear combination of species mass fractions,
\begin{equation}
    \mathcal{Y} = \sum_{i=1}^{N_s} c_{\mathcal{Y},i} Y_i\; .
\end{equation}
Here is $c_{\mathcal{Y},i}$ the progress variable coefficient of species $i$. In this work, the progress variable is constructed solely of the species H$_2$ and H$_2$O both weighted by their corresponding molar masses. This results in a monotonically increasing progress variable, which is essential to ensure a unique lookup point in the FGM database. The progress variable coefficients are in $c_{\mathcal{Y},H_2} = -1 $ and $c_{\mathcal{Y},H_2O} = M_{H_2}/M_{H_2O}$, while $c_{\mathcal{Y},i} = 0 $ for the remaining species. 

Since the Lewis number of hydrogen is significantly lower than unity, preferential diffusion effects result in local changes in mixture composition. To address this, a second control variable $\phi_2$ is introduced: the element based mixture fraction of the hydrogen element,
\begin{equation}
    Z = \sum_{i=1}^{N_s} c_{Z,i} Y_i\; ,
\end{equation}
with $c_{Z,i}$ being the mass fraction of the hydrogen element in species $i$. In most engineering applications, heat losses play an important role. Therefore, a third control variable $\phi_3$ is introduced: the enthalpy,
\begin{equation}
    h = \sum_{i=1}^{N_s} c_{h,i}(T) Y_i\; ,
\end{equation}
where $c_{h,i}(T)$ is the specific enthalpy of species $i$, earlier defined as $h_i$. 
\subsection{Preferential Diffusion Analysis}
\label{sec:preferentialdiffusion}
\begin{figure}
     \centering
     \begin{subfigure}[b]{0.49\textwidth}
         \centering
         \includegraphics[width=\textwidth]{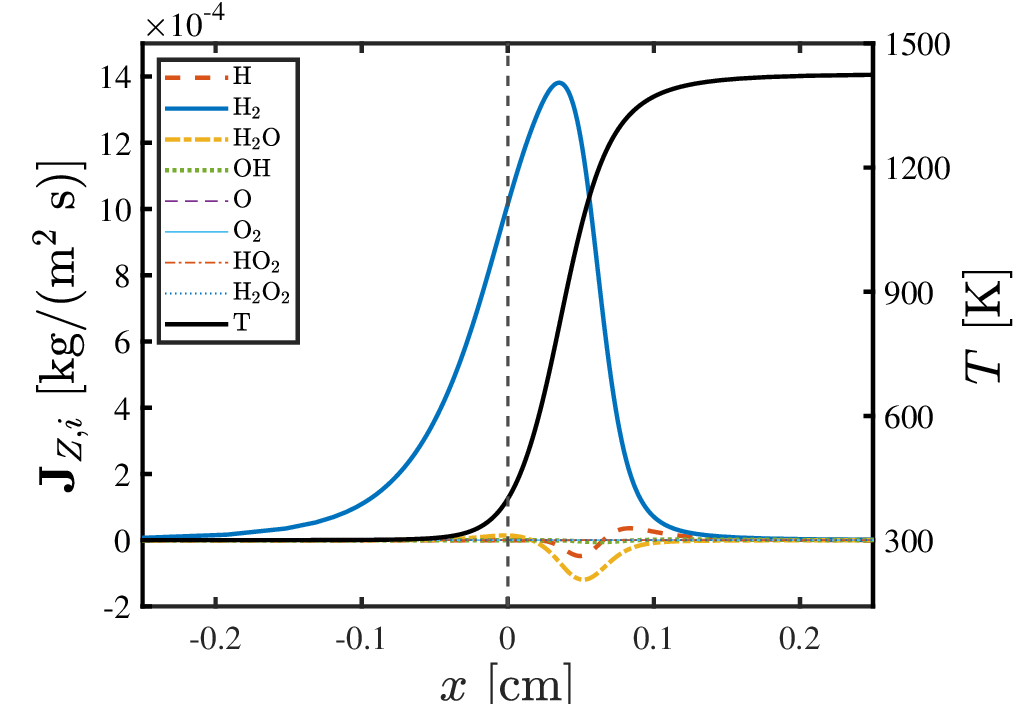}
         \caption{$\phi = 0.4$}
         \label{fig:prefdiffcontributionsZ}
     \end{subfigure}
     \hfill
     \begin{subfigure}[b]{0.49\textwidth}
         \centering
         \includegraphics[width=\textwidth]{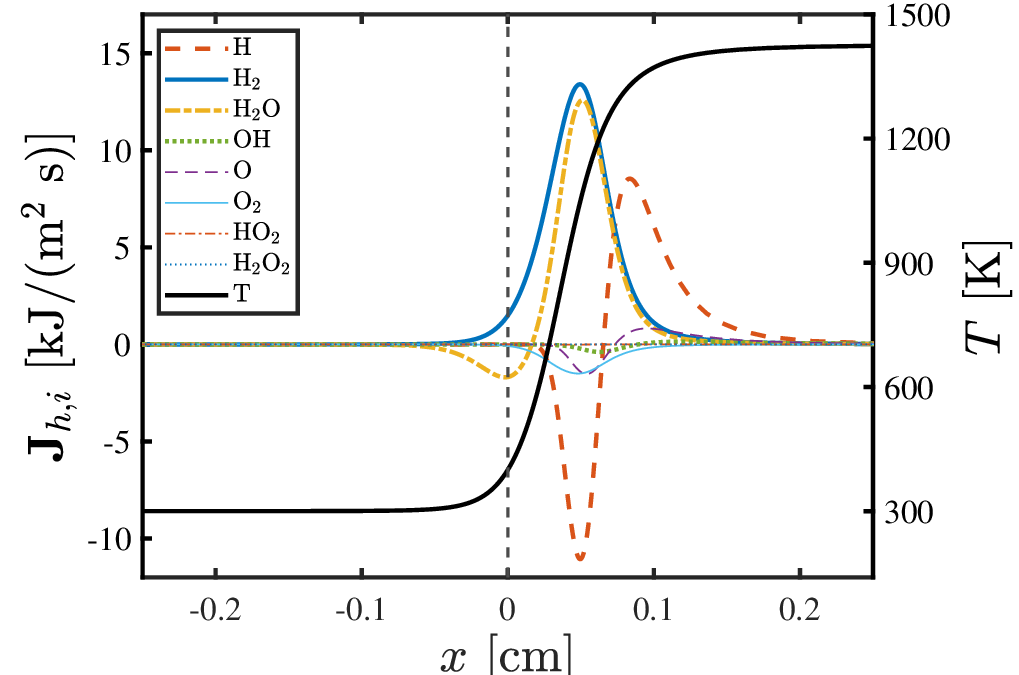}
         \caption{$\phi = 0.4$}
         \label{fig:prefdiffcontributionsh}
     \end{subfigure}
     \begin{subfigure}[b]{0.49\textwidth}
         \centering
         \includegraphics[width=\textwidth]{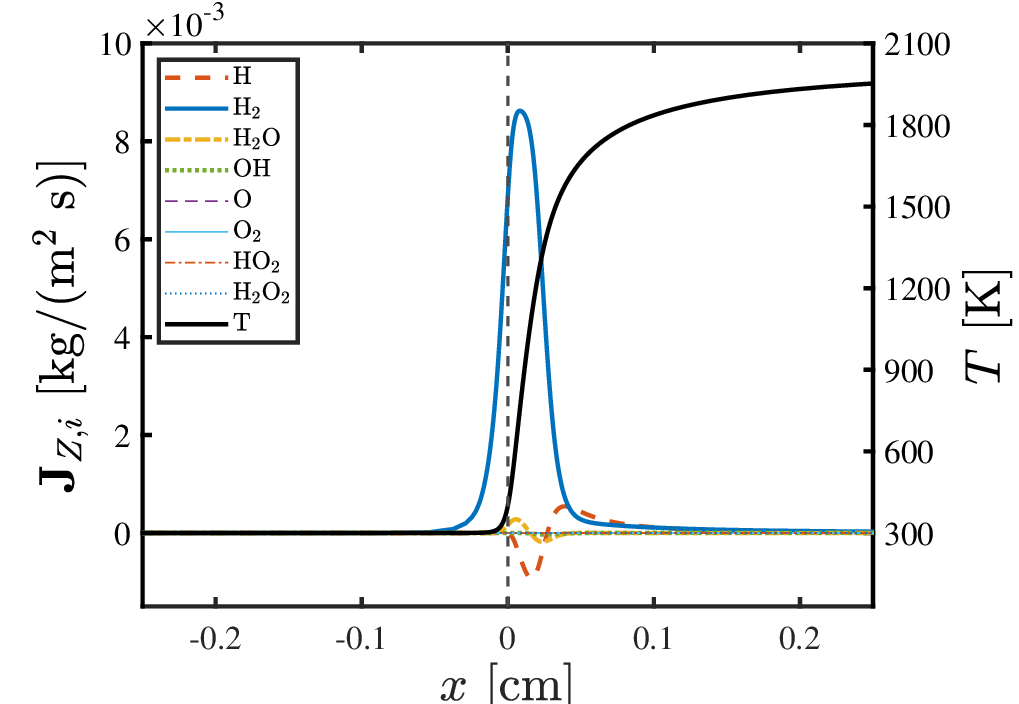}
         \caption{$\phi = 0.7$}
         \label{fig:prefdiffcontributionsZ_07}
     \end{subfigure}
     \hfill
     \begin{subfigure}[b]{0.49\textwidth}
         \centering
         \includegraphics[width=\textwidth]{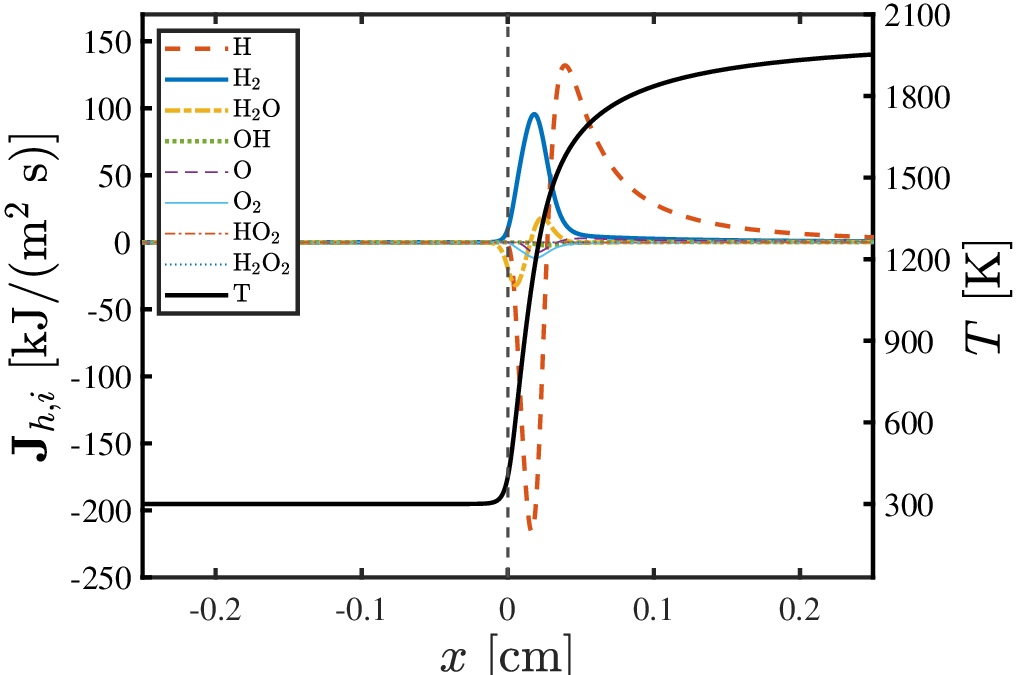}
         \caption{$\phi = 0.7$}
         \label{fig:prefdiffcontributionsh_07}
     \end{subfigure}
     
     \caption{The contributions of each chemical species to the preferential diffusion flux of mixture fraction (a,c) and enthalpy (b,d), and the temperature profile obtained from a one-dimensional premixed hydrogen-air flame simulation at an equivalence ratio of $\phi = 0.4$ (a,b) and $\phi = 0.7$ (c,d).}
     \label{fig:prefdiffcontributions}
\end{figure}

The preferential diffusion term in Eq. (\ref{eq:controlvariablediffusionflux}) can be computed by storing the diffusion coefficients and mass fractions of all chemical species present in the chemical mechanism in the lookup table. During the CFD calculation, these coefficients and mass fractions are retrieved using the control variables, and the diffusion fluxes of all species are then calculated to construct the preferential diffusion term. However, for large chemical mechanisms this results in a significant increase in memory occupation of the lookup table and extra computational overhead due to retrieving all variables and computing the species gradients. A simpler method could be to include solely the major species which have a significant contribution to the total preferential diffusion flux of the control variable, which are species that either have $\text{Le} \neq 1$ or a large thermal diffusion coefficient $D_i^T$. To analyse this, the contributions of each chemical species to the preferential diffusion flux of the mixture fraction and enthalpy, i.e., the terms within the summation of the right term in Eq. (\ref{eq:controlvariablediffusionflux}), are plotted in Figure \ref{fig:prefdiffcontributions}. The results are obtained from one-dimensional flame simulations at ambient pressure ($101325$ Pa) and temperature ($300$ K).  The mixture consists of hydrogen in air at two equivalence ratios: $\phi = 0.4$ and $0.7$.  These cases are included to illustrate that the relative flux contributions of individual species remain similar across the range of equivalence ratios that can occur due to thermo-diffusive instabilities. The chemical mechanism used is by Burke et al.~\cite{Burke2011ComprehensiveCombustion} and the simulations are performed using an in-house one-dimensional flamelet code Chem1D \cite{70136d09e89b41be89fb05427ca596b8}. 

In all plots of Figure \ref{fig:prefdiffcontributions}, the dotted vertical line indicates the point where the temperature is equal to 400 K, defined as the zero-position. In Figure \ref{fig:prefdiffcontributionsZ} and \ref{fig:prefdiffcontributionsZ_07}, it can be observed that there is a positive flux of mixture fraction due to the diffusion of hydrogen before zero-position, which is a result of hydrogen's low Lewis number. From Figures \ref{fig:prefdiffcontributionsZ} and \ref{fig:prefdiffcontributionsZ_07}, it can be concluded that the preferential diffusion flux of the mixture fraction is predominantly governed by H$_2$, contributing approximately 94\% at $\phi = 0.4$ and 93\% at $\phi = 0.7$. The contributions of H$_2$O and H are significantly smaller: around 4\% and 1\%, respectively, at $\phi = 0.4$, and 2\% and 5\% at $\phi = 0.4$. All other species seem to have no significant influence on the preferential diffusion of mixture fraction. The species contributions to the preferential diffusion of enthalpy, plotted in Figure \ref{fig:prefdiffcontributionsh} and \ref{fig:prefdiffcontributionsh_07}, seem to be more diverse. The species H, H$_2$ and H$_2$O seem to be the most significantly contributing species ($\approx 90 \%$ for both $\phi$), while species such as O$_2$, O and OH having a slight contribution ($\approx 7 \%$) as well. The progress variable is excluded from this analysis as it is constructed solely from H$_2$ and H$_2$O, which results in $\mathbf{J}_{\mathcal{Y},i} = 0$ except for  H$_2$ and H$_2$O.

\subsection{Manifolds \& Lookup}
\label{sec:manifolds}

Two manifolds are constructed: Manifold A and Manifold B. Manifold A is a two-dimensional manifold constructed by simulating premixed flamelets over a range of inlet equivalence ratios $\phi$. This manifold is parameterized by the progress variable and mixture fraction and should therefore be able to capture the reaction progress as well as local changes in mixture composition due to preferential diffusion. Manifold B is a three-dimensional manifold which is constructed by simulating premixed flamelets over a range of inlet equivalence ratios $\phi$ and a range of inlet enthalpy values $h_{\text{in}}$. This manifold is parameterized by a progress variable, mixture fraction and enthalpy, and should therefore be able to capture heat loss effects in addition to the effects captured by manifold A. Based on these two manifolds, three different models are created which have a varying set of major species contributing to the preferential diffusion of the control variables. These models are elaborated separately below.\\

\noindent\textbf{FGM model A}\\

The first model, FGM A, is based upon a Manifold A. It is assumed that the preferential diffusion flux of mixture fraction will solely be governed by the major species H$_2$ and H$_2$O. The contribution of the remaining species, which is mostly H, is gathered in a closure source term $S_{\phi_k}$. For mathematical consistency, the diffusion flux of the control variables, found in Eq. (\ref{eq:controlvariablediffusionflux}), is rewritten by introducing a coefficient $\Lambda_i$, which is in this case defined as
\begin{equation}
\label{eq:lambda_i_1}
    \Lambda_i = \begin{cases}
      1 & \text{if} \;i \in \left\{\text{H}_2, \text{H}_2\text{O} \right\} \\
      0 & \text{if} \;i \notin \left\{\text{H}_2, \text{H}_2\text{O} \right\}
    \end{cases}
    \;, \qquad i\in[1,N_s] \;.
\end{equation}
This coefficient is incorporated in Eq. (\ref{eq:controlvariablediffusionflux}) and the species summation is split in a molecular and thermal diffusion term: 
\begin{equation}
\label{eq:controlvariableDiffusionA}
\begin{split}
    \mathbf{J}_{\phi_k} = - \sum_{i=1}^{N_s} \Lambda_i  \underbrace{c_{\phi_{k,i}}\left(\rho D_i - \frac{\lambda}{c_p} \right)}_{D_{\phi_{k,i}}}\nabla Y_i - \underbrace{\left(\frac{\rho}{T}\sum_{i=1}^{N_s} \Lambda_i c_{\phi_{k,i}} D_i^T Y_i \right)}_{D_{\phi_k}^T}\nabla T\; .
\end{split}
\end{equation}
Splitting the summation allows for defining a set of molecular diffusion coefficients $D_{\phi_k,i}$ for $i \in \left\{\text{H}_2, \text{H}_2\text{O} \right\}$, and a single thermal diffusion coefficient $D_{\phi_k}^T$. As mentioned, the contribution of the remaining species is gathered in the diffusion closure source term $S_{\phi_k}$, which is added to the control variable transport equation:
\begin{equation}
    \label{eq:controlvariableconservationA}
    \frac{\partial\left(\rho \phi_k\right)}{\partial t} + \nabla\cdot\left(\rho \mathbf{u} \phi_k\right) - \nabla\cdot\left(\frac{\lambda}{c_p}\nabla \phi_k\right) = -\nabla\cdot \mathbf{J}_{\phi_k} + S_{\phi_k} + \dot{\omega}_{\phi_k}\;, \qquad k\in[1,N_{c}] \;.
\end{equation}
Because we compute $S_{\phi_k}$ during pre-processing, the divergence of the one-dimensional diffusion flux can be written as a derivative with respect to the flamelet coordinate $x$, which is not the same as the $x$ coordinate in the multi-dimensional CFD simulation. The diffusion closure source term $S_{\phi_k}$ can thus be written as
\begin{equation}
\label{eq:controlvariableclosure}
    S_{\phi_k} = \frac{d}{dx} \sum_{i=1}^{N_s} (1 - \Lambda_i )c_{\phi_{k,i}} \left[\left(\rho D_i - \frac{\lambda}{c_p}\right) \frac{dY_i}{dx} - \frac{\rho D_i^TY_i}{T}\frac{dT}{dx}\right]\;.
\end{equation}
In this way, we attempt to approximate the three-dimensional divergence of the diffusion flux of the remaining species using the one-dimensional divergence in the flamelets. This introduces an error, but following Figure \ref{fig:prefdiffcontributions}, it will be assumed to be negligible.

The preferential diffusion coefficients, $D_{\phi_k,i}$ for $i \in \left\{\text{H}_2, \text{H}_2\text{O} \right\}$ and $D_{\phi_k}^T$, along with the mass fractions of H$_2$ and H$_2$O, as well as the diffusion closure source term $S_{\phi_k}$, are pre-computed, stored in the manifold, and retrieved during the calculation. The hypothesis for FGM model A is that the TD instabilities due to preferential diffusion effects can accurately be modeled by only incorporating the contributions of H$_2$ and H$_2$O directly in the preferential diffusion of the control variables, i.e., progress variable and mixture fraction, and gathering the small remaining contributions of other species in a source term.\\

\noindent\textbf{FGM model B1}\\

The second model, FGM B1, is based upon a Manifold B. The preferential diffusion is again assumed to be solely governed by the major species H$_2$ and H$_2$O. The coefficient $\Lambda_i$ remains the same as for manifold A as given in Eq. (\ref{eq:lambda_i_1}), and the general control variable equations in Eqs. (\ref{eq:controlvariableDiffusionA}), (\ref{eq:controlvariableconservationA}) and (\ref{eq:controlvariableclosure}) still hold. Note that the significant contribution of the H-radical, which followed from Figure \ref{fig:prefdiffcontributionsh}, is not incorporated directly in the preferential diffusion term, but in the source term $S_{\phi_k}$. This model is similar to the approach taken by Böttler et al.~\cite{Bottler2023FlameletFlames}. In their model, they solve transport for a set of major species, H$_2$, H$_2$O and O$_2$, and enthalpy, construct a progress variable and mixture fraction from the set of major species during the CFD simulation, and use the fluxes of the major species in the preferential diffusion term of the enthalpy. Using this method, the contribution of the H-radical is also not directly incorporated in the preferential diffusion of enthalpy, but again through the closure source term, which in their work is referred to as $\nabla h_{\text{closure}}$. By doing this, transport along the flame front (cross-diffusion) due to a flux of H is not incorporated, while from Figure \ref{fig:prefdiffcontributionsh} it was concluded that H does have a significant contribution to the preferential diffusion of enthalpy. The hypothesis is that by excluding the contribution of the H-radical to the preferential diffusion of enthalpy directly, the cross-diffusion and TD instabilities due to preferential diffusion and curvature are not accurately captured by FGM model B1.\\

\noindent\textbf{FGM model B2}\\

The third model, FGM B2, is also based upon a Manifold B. This time, the preferential diffusion is assumed to be governed by the major species H$_2$, H$_2$O and H. This results in a different coefficient,
\begin{equation}
\label{eq:lambda_i_2}
    \Lambda_i^* = \begin{cases}
      1 & \text{if} \;i \in \left\{\text{H}, \text{H}_2, \text{H}_2\text{O} \right\} \\
      0 & \text{if} \;i \notin \left\{\text{H}, \text{H}_2, \text{H}_2\text{O} \right\}
    \end{cases}
    \;, \qquad i\in[1,N_s] \;.
\end{equation}
Note that the small contribution of the H-radical to the mixture fraction preferential diffusion, which can be seen in Figure \ref{fig:prefdiffcontributionsZ} and \ref{fig:prefdiffcontributionsZ_07}, is now also included directly in the preferential diffusion term. The small contributions of the remaining species to the preferential diffusion of enthalpy, mainly O$_2$, O and OH as seen in Figure \ref{fig:prefdiffcontributionsh} and \ref{fig:prefdiffcontributionsh_07}, are gathered in the closure source term $S_h$. Note that for the progress variable, the contribution of the H-radical is zero, since the progress variable is constructed from H$_2$ and H$_2$O only. The hypothesis is that by including the contribution of the H-radical directly in the preferential diffusion of enthalpy, cross-diffusion and the TD instabilities due to preferential diffusion and curvature can be modeled with significantly more accuracy using FGM B2 compared to FGM B1.\\
An overview of the FGM models considered in this work is given in Table \ref{tab:manifolds}.
\begin{table}
\centering
\caption{This table presents an overview of the manifolds considered in this work; their control variables and the major species selection contributing to the preferential diffusion term}
\renewcommand{\arraystretch}{1.1}

\begin{tabular}{lll}
\toprule
FGM model & Control variables & Major species selection \\ \midrule
A & $\mathcal{Y}$, $Z$ & H$_2$, H$_2$O \\ 
B1 & $\mathcal{Y}$, $Z$, $h$ & H$_2$, H$_2$O \\
B2 & $\mathcal{Y}$, $Z$, $h$ & H$_2$, H$_2$O, H \\
\bottomrule
\end{tabular}
\label{tab:manifolds}
\end{table}

The FGM chemistry model including the control variable equations and the lookup routine is incorporated in the compressible OpenFOAM solver, \textit{rhoPimpleFoam}. The table is constructed by generating a set of flamelets using the one-dimensional flame solver Chem1D \cite{70136d09e89b41be89fb05427ca596b8} and interpolating the data on a sparse rectilinear grid with 250 points in $\mathcal{Y}$, 100 points in $Z$ and 100 points in $h$.

\section{Results and Verification}
\label{sec:results_and_verification}

The FGM models presented in Section \ref{sec:numerical_model} are analyzed by simulating a planar laminar TD unstable lean hydrogen-air flame, and comparing the results to a reference detailed chemistry (DC) simulation. First, the case setup and simulation settings are detailed, followed by an \textit{a-priori} comparison of the FGM models to DC. Next, a linear stability analysis is performed, and finally, the \textit{a-posteriori} results of the unstable flame front are presented. 

\subsection{Setup}
\label{sec:setup}

The simulation setup is similar to the work of Berger et al.~\cite{Berger2019CharacteristicFlames}. The two-dimensional domain has a height $L_y = 200l_F$, where $l_F$ represents the unstretched laminar flame thickness defined as $l_F=\left(T_b - T_u\right)/\max\left(\frac{dT}{dx}\right)$. In the work of Berger et al.~\cite{Berger2019CharacteristicFlames}, the domain width $L_x$ is varied and it is shown that the average flame consumption speed stabilizes for $\log_{10}(L_x/l_F) > 2$, which corresponds to a domain width of $L_x > 100l_F$. Therefore, a domain width of $100l_F$ is chosen. The mesh resolution was set to 20 cells in the unstretched laminar flame front, i.e. $\Delta x = l_F/20$. The authors found that further refinement beyond this resolution did not affect the dynamic behavior for any of the models considered. The initial flame front is perturbed with a sinusoidal disturbance with a wavelength $\lambda=10l_F$. The considered fuel mixture is hydrogen in air at an equivalence ratio $\phi = 0.4$ at an ambient pressure of $101325$ Pa and with a temperature of $300$ K. At these conditions, $l_F$ has a value of approximately 0.67 mm. A uniform inlet velocity is chosen to balance the flame propagation speed and keep the flame front in the computational domain. The velocity has a value of $0.70$ m/s and is kept constant for all simulated cases. For both the detailed chemistry calculation and the flamelets for FGM table generation, the chemical mechanism by Burke et al.~\cite{Burke2011ComprehensiveCombustion} is used. All simulations run with mixture averaged diffusion coefficients and thermal (Soret) diffusion. The total simulation time for all cases is approximately 150 $\tau_F$, where $\tau_F$ is the laminar flame time, defined as $\tau_F=l_F/s_L$. Here, $s_L$ denotes the unstretched laminar flame speed, which has a value of 204.85 mm/s at these conditions, resulting in $\tau_F = 3.26$ ms. Note that the inlet velocity is more than 3 times higher than the laminar burning velocity of this mixture. This is because the increased flame surface area due to the unstable wrinkled flame front results in a much larger burning rate. The FGM model's computational cost is about an order of magnitude higher than that of detailed chemistry (DC), while the difference between individual FGM models is significantly smaller.

\subsection{A-priori Validation of the Manifolds}
\label{sec:apriori}

\begin{figure}
    \centering
    \includegraphics[width=0.7\textwidth]{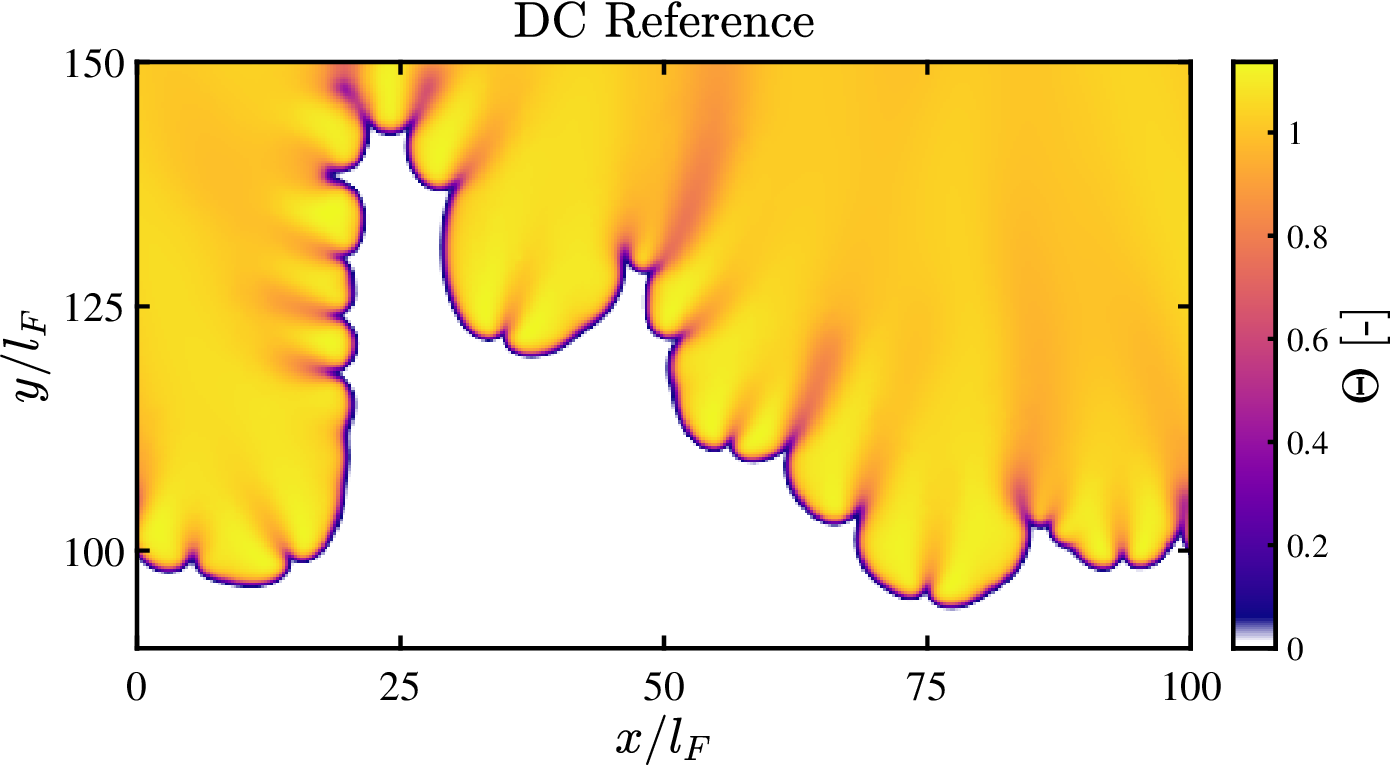}
    \caption{Contour plot of the normalized temperature $\Theta = (T - T_\mathrm{u})/(T_\mathrm{ad} - T_\mathrm{u})$ for the detailed chemistry (DC) reference case.}
    \label{fig:temperatureContourDC}
\end{figure}

Before simulations with the FGM models are performed, the manifolds A and B are analysed in an a-priori manner. A reference case is simulated in accordance with the setup described in Section \ref{sec:setup} using the detailed chemistry (DC) model. As described in Section \ref{sec:reactiveflows}, a mixture averaged diffusion model including thermal (Soret) diffusion is used. Additionally, a correction velocity is applied to the species equations to ensure global mass conservation, as described by Poinsot and Veynante \cite{ThierryPoinsot2005TheoreticalCombustion}. Figure \ref{fig:temperatureContourDC} shows the normalized temperature contour at $\tau_F \approx 70$. The characteristic flame tips, similar to those reported in several literature sources \cite{Altantzis2011DetailedFlames,Altantzis2012HydrodynamicFlames,Kadowaki2005TheInstability,Berger2019CharacteristicFlames,Berger2023FlameFlames,Creta2020PropagationInstabilities,Kadowaki2005NumericalInteraction,Wen2024ThermodiffusivelyPatterns}, are visible. Further analysis of these characteristic patterns and their dynamical evolution will follow in Section \ref{sec:simulationresults}.

\begin{figure}
    \centering
    \includegraphics[width=\textwidth]{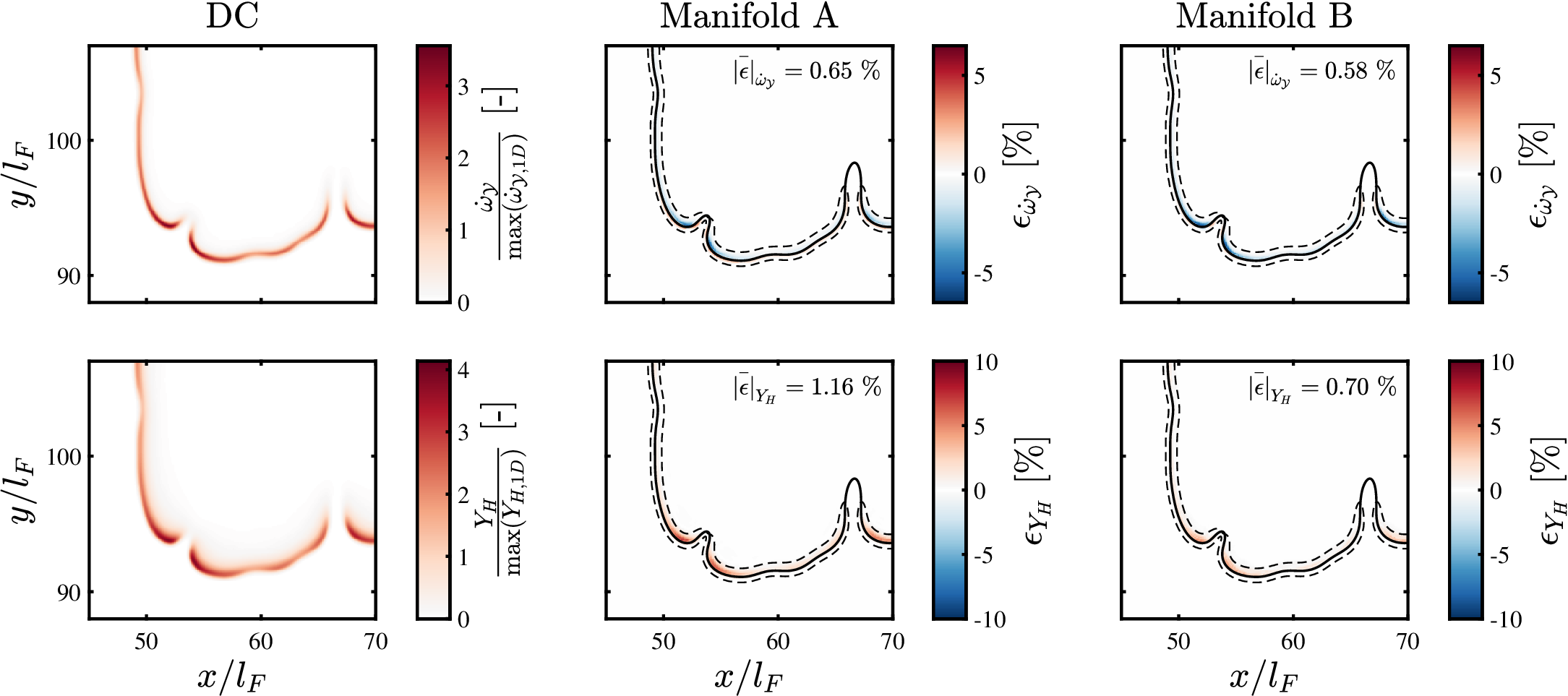}
    \caption{A-priori analysis of manifold A and manifold B using the DC data. The left column show DC values normalized by their 1D maximum values. The right two columns show the normalized error of the manifolds compared to the DC values, defined as $\epsilon_\psi = (\psi_{\text{DC}}-\psi_{\text{FGM}})/\text{max}(\psi_{\text{DC}}) \cdot 100\%$.}
    \label{fig:apriori}
\end{figure}

In Figure \ref{fig:apriori} the \textit{a-priori} analysis of the manifolds is visualized. The normalized progress variable source term and the normalized H-radical mass fraction obtained from the DC reference calculation are plotted in the upper left and lower left figures, respectively. The control variables associated with each manifold are obtained from the DC fields and a lookup of the variables in consideration is performed. The difference between the manifold lookup variables and the DC values is quantified by the normalized error $\epsilon_\psi$. The magnitude of this error is integrated over and divided by the area enclosed by the dashed lines, which defined as where the normalized progress variable source term is higher than a threshold value of $\frac{\dot{\omega}_\mathcal{Y}}{\text{max}(\dot{\omega}_{\mathcal{Y},\text{1D}})} = 0.165$. This average error value is added to the plot as $|\bar{\epsilon}|_\psi$. It can be seen that by adding enthalpy as an extra dimension to the manifold, both variables are represented with slightly higher accuracy. The contour plot of the progress variable source term error $\epsilon_{\dot{\omega}_\mathcal{Y}}$ does not seem to show a significant improvement, but the average error magnitude $|\bar{\epsilon}|_{\dot{\omega}_\mathcal{Y}}$ does show a slight improvement of manifold B with respect to manifold A\@. The surface plots of the H-radical error $\epsilon_{Y_H}$ seem to show a slightly more significant error reduction in the flame tip region, which is also reflected by a reduced average error magnitude $|\bar{\epsilon}|_{\dot{\omega}_\mathcal{Y}}$. Nonetheless, it can be concluded that both manifolds show no significant deviations from the DC reference case when compared in an \textit{a-priori} sense.

\subsection{Linear Stability Analysis}
\begin{figure}
    \centering
    \includegraphics[width=0.5\textwidth]{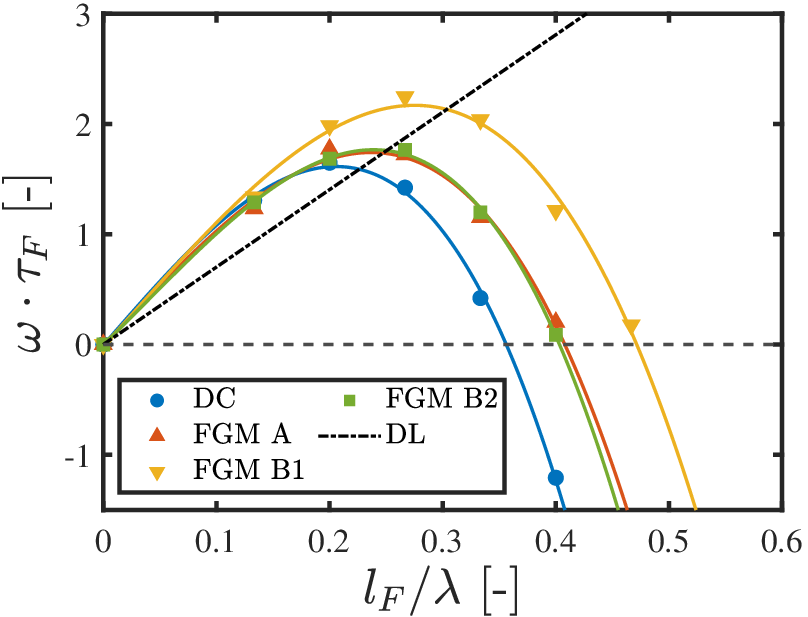}
    \caption{A comparison of the non-dimensional dispersion relation between detailed chemistry (DC) and the FGM models. The dots represent the calculated dimensionless growth rates and the lines represent a third order polynomial fit through these points. The black dash-dotted line represents the Darrieus-Landau growth rate obtained from the work of Matalon \cite{Matalon2018}.}
    \label{fig:LSA}
\end{figure}

The exponential growth of small perturbations ultimately leads to the typical cell structures found in TD unstable flames. Therefore, the growth of perturbations for different wave lengths is studied by performing a linear stability analysis (LSA), which has proven to be a useful tool in investigating flame front instabilities \cite{Berger2019CharacteristicFlames,Berger2023FlameFlames}. The planar flame front is disturbed with a small initial perturbation of $A_0=0.04l_F$ for varying wave lengths $\lambda$, similar as in the work of Berger et al.~\cite{Berger2019CharacteristicFlames} and Böttler et al.~\cite{Bottler2023FlameletFlames}. The lateral domain size is adjusted such that exactly one wave length fits in the domain, and to ensure accurate results, a high mesh resolution of 100 cells in the laminar flame front is used, i.e. $\Delta x=l_F/100$. For each wave length, the growth rate $\omega = d(\ln{A(t)}/A_0)/dt$ is calculated. Figure \ref{fig:LSA} shows the dimensionless growth rate for varying dimensionless wave numbers. A positive number indicates an unstable exponential growth of the perturbation, while a negative number indicates an exponential dampening of the perturbation.\\
The global behavior seems to be consistent, showing a growth of perturbations when the wavelengths are not too small, and a peak around $4l_F$. Additionally, all FGM models show unstable growth of perturbations at smaller wavelengths compared to the DC reference case. However, FGM model B1 seems to deviate the most from DC, showing similar results to the FPV-$h$ model presented in the work of Böttler et al.~\cite{Bottler2023FlameletFlames}. This can be due to the fact that these models are in their core very similar. The FPV-$h$ model does not explicitly take into account the contribution of the H-radical to the preferential diffusion of enthalpy, since their major species selection only consists of H$_2$, O$_2$ and H$_2$O. By explicitly including the H-radical in the enthalpy preferential diffusion term, as done for FGM model B2, a significant improvement can be observed. FGM B2 also shows a slight improvement compared to the FPV-$\kappa_c$ model proposed by Böttler et al.~\cite{Bottler2023FlameletFlames}, which attempts to include curvature effects in the manifold. FGM A also appears to perform well, which is expected given the small errors observed in the \text{a-priori} analysis in Section \ref{sec:apriori}. This indicates that the main deviations arise not from the manifolds themselves but from the solution of the control variable transport equations. Specifically, our analysis suggests that discrepancies in the predicted control variables, mainly influenced by the preferential diffusion formulation as well as the general FGM assumption, play a significant role in altering the flame's stability characteristics. To further analyze the models, an unstable flame front is simulated for a longer time to study the subsequent non-linear behavior.

\subsection{A-posteriori Unstable Flame Front Simulation}
\label{sec:simulationresults}

\begin{figure}
    \centering
    \includegraphics[width=\textwidth]{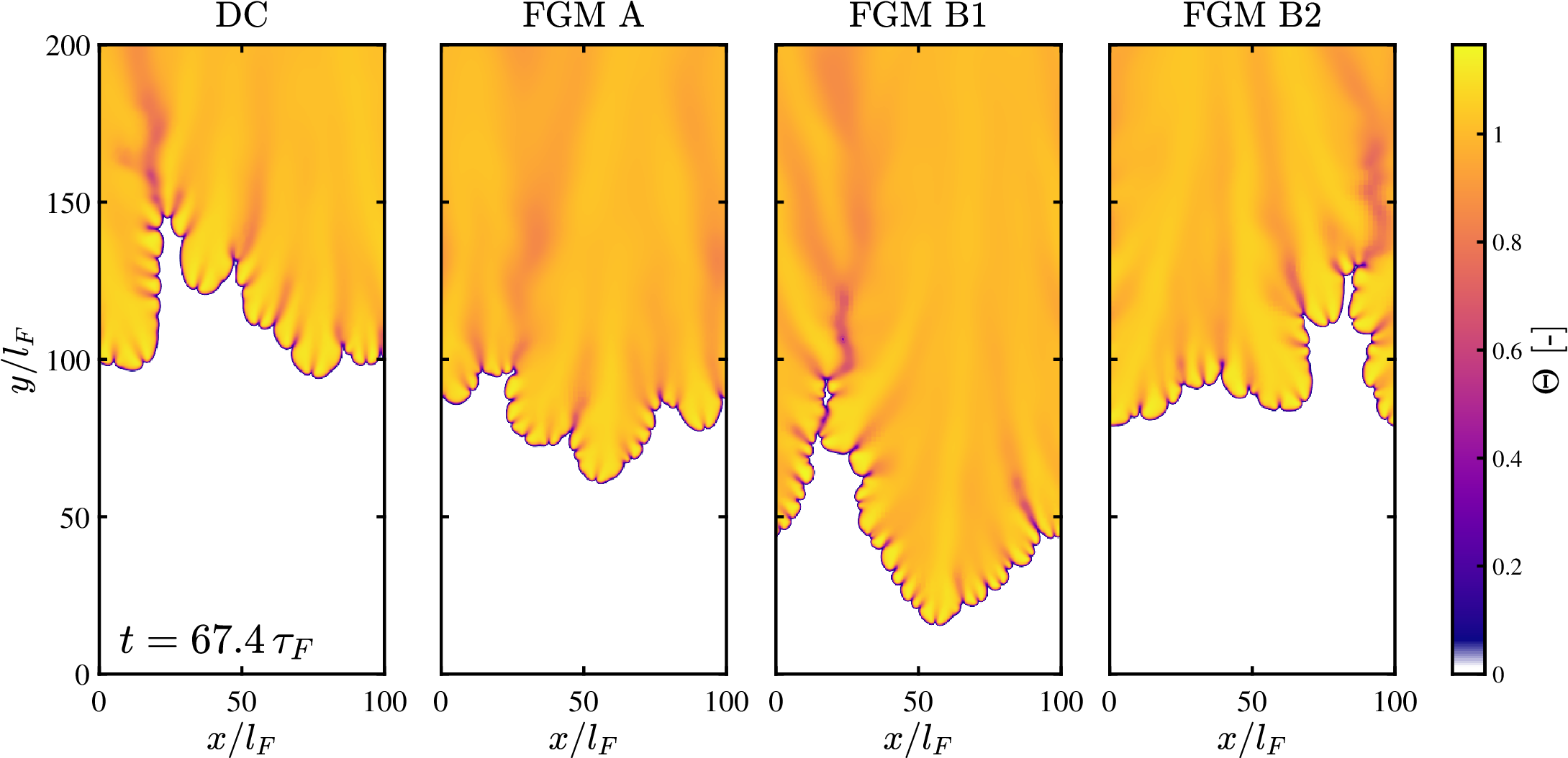}
    \caption{Contour plot of the normalized temperature $\Theta = (T - T_\mathrm{u})/(T_\mathrm{ad} - T_\mathrm{u})$ for the compared models.}
    \label{fig:temperatureContour}
\end{figure}

In Figure \ref{fig:temperatureContour}, the DC calculation is compared with the FGM models by lining up snapshots of the normalized temperature contours at $t \approx 67.4\tau_F$. The temperature is normalized using unburnt temperature $T_{\text{u}}$ of 300 K and the adiabatic flame temperature $T_{\text{ad}}$ of the inlet mixture ($\phi = 4.0$). It can be observed that in post-flame regions where the flame is positively curved, i.e., convex towards the incoming flow, the normalized temperature increases to values above unity. Due to preferential diffusion, relatively more H$_2$ diffuses toward these regions, leading to a richer mixture and consequently a super-adiabatic temperature. In contrast, in negatively curved regions, i.e., concave towards the incoming flow, the mixture becomes leaner, leading to a trail of sub-adiabatic temperatures. This effect is visible in all models. While investigating the shape of the wrinkled flame front, DC, FGM model A and B2 appear to show the characteristic flame tips, similar to those reported in several literature sources \cite{Altantzis2011DetailedFlames,Altantzis2012HydrodynamicFlames,Kadowaki2005TheInstability,Berger2019CharacteristicFlames,Berger2023FlameFlames,Creta2020PropagationInstabilities,Kadowaki2005NumericalInteraction,Wen2024ThermodiffusivelyPatterns}. These flame tips consist of a single flame cusp enclosed by two flame bulges. This structure remains stable and tilts towards an incline with respect to the incoming flow. This results in a lateral displacement in positive or negative $x$-direction depending on the flame tip orientation, until the flame tip reaches a different part of the flame front and merges. FGM model B1 deviates from this dynamic behavior. The characteristic flame tip structure is still present, but it does not exhibit the dynamic instability or lateral tilting observed in the other models. Instead, a single structure remains stable, oriented perpendicular to the flow, and propagates upstream, while smaller secondary flame tips are suppressed and rarely emerge.

\begin{figure}
    \centering
    \includegraphics[width=\textwidth]{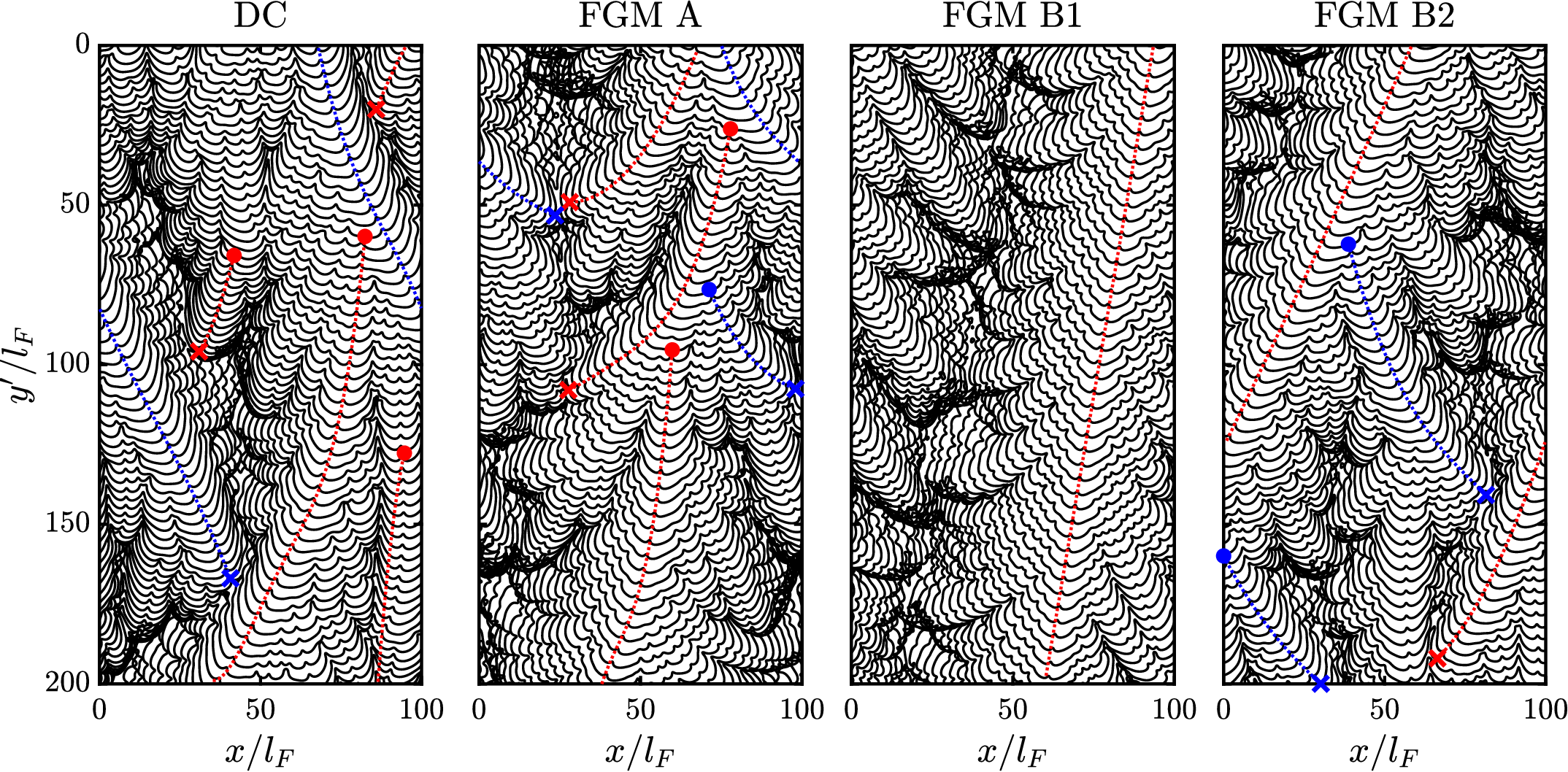}
    \caption{Flame front visualisation to analyse the dynamic behaviour of the different models. The solid black lines indicate the flame front at different time instances, defined as the  $\Theta = 0.5$  normalized temperature isocontour. The colored dashed lines show the movement of the flame cusps for all the models.}
    \label{fig:flamecusptracking}
\end{figure}

To further highlight the differences in dynamic behavior between the models, snapshots of the flame fronts at several time instances are plotted in Figure \ref{fig:flamecusptracking}. The black lines represent the flame fronts, defined as the $\Theta = 0.5$ isocontour. The flame front that enters the figure at $y'/l_F = 0$ corresponds to $t = 38\tau_F$. For each time instance the flame front is displaced downwards with a value $\Delta y' = U\Delta t$, where $U$ represents the inflow velocity and $\Delta t \approx 0.613\tau_F$. This is done to facilitate a clear distinction between the flame fronts. Some of the flame cusps for all the models, are tracked by the dashed blue and red lines which represent the flame cusps moving in positive and negative x-direction, respectively. The colored dots and crosses represent the locations of emergence and disappearance of the flame cusps, respectively. 

For DC, FGM A and FGM B2, the flame tips that enter at $y'/l_F = 0$ are either already inclined, or proceed to tilt towards an angle with respect to the incoming flow. New flame cusps seem to emerge from the existing laterally displacing flame tips, where two bulges on each side expand, resulting in a new flame tip structure. The flame tips get destroyed by collision with other parts of the flame front due to their lateral displacement. This behavior is in line with the findings of Berger et al.~\cite{Berger2019CharacteristicFlames}. In contrast to the other models, FGM B1 does not show this behavior. A single flame finger enters the domain and continues to propagate downstream while maintaining a stable orientation perpendicular to the incoming flow. Although smaller secondary flame tips occasionally emerge from the sides of this dominant structure, they are destroyed upon interaction with the main flame tip. Due to this persistent dominant flame tip, the flame front moves downwards significantly faster compared to the other models.



\begin{figure}
    \centering
    \includegraphics[width=0.5\textwidth]{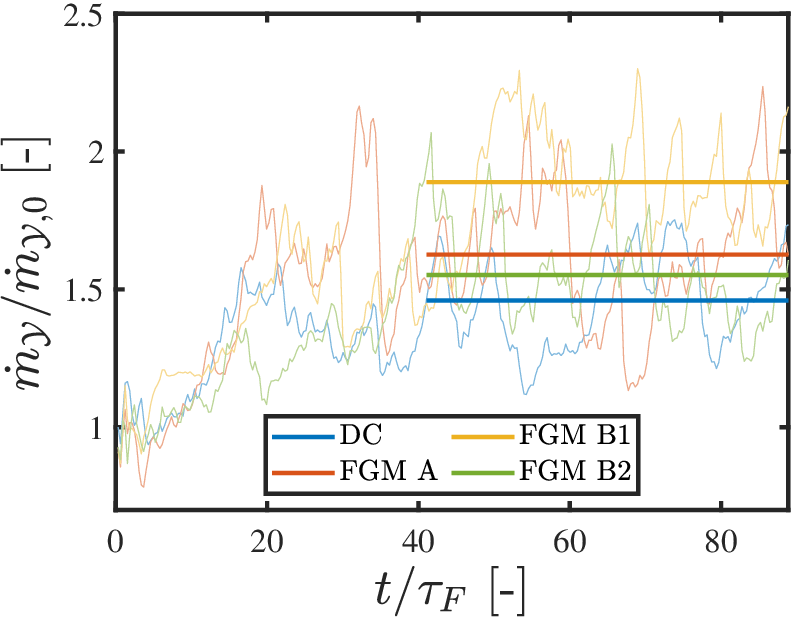}
    \caption{The progress variable mass production rate normalized by the initial unstretched value. The thin colored lines indicate the instantaneous values and the thick colored lines the time averaged values.}
    \label{fig:PVmassconsumptionrate}
\end{figure}
This is also evident when looking at the total progress variable chemical production rate over the simulation time in Figure \ref{fig:PVmassconsumptionrate}. The chemical production rate is computed by integrating the progress variable source term over the domain. The thin lines represent the instantaneous values and the thick lines the averaged values over the window $t/\tau_F \in [40,85]$. An averaging start time of $t/\tau_F = 40$ is selected to ensure that initialization effects do not affect the averaged values.

One can see that the averaged progress variable production rates in Figure \ref{fig:PVmassconsumptionrate} of the DC, FGM A and FGM B2 are significantly lower than of FGM B1. Because the inflow velocity is kept constant for all simulations, this results in the flame propagating much faster towards the inlet for FGM B1. Notably, model FGM A also exhibits a slight overprediction in burning rate, suggesting that the enthalpy itself may contribute to an accurate prediction of thermo-diffusive instabilities.



\begin{figure}
    \centering
    \includegraphics[width=\textwidth]{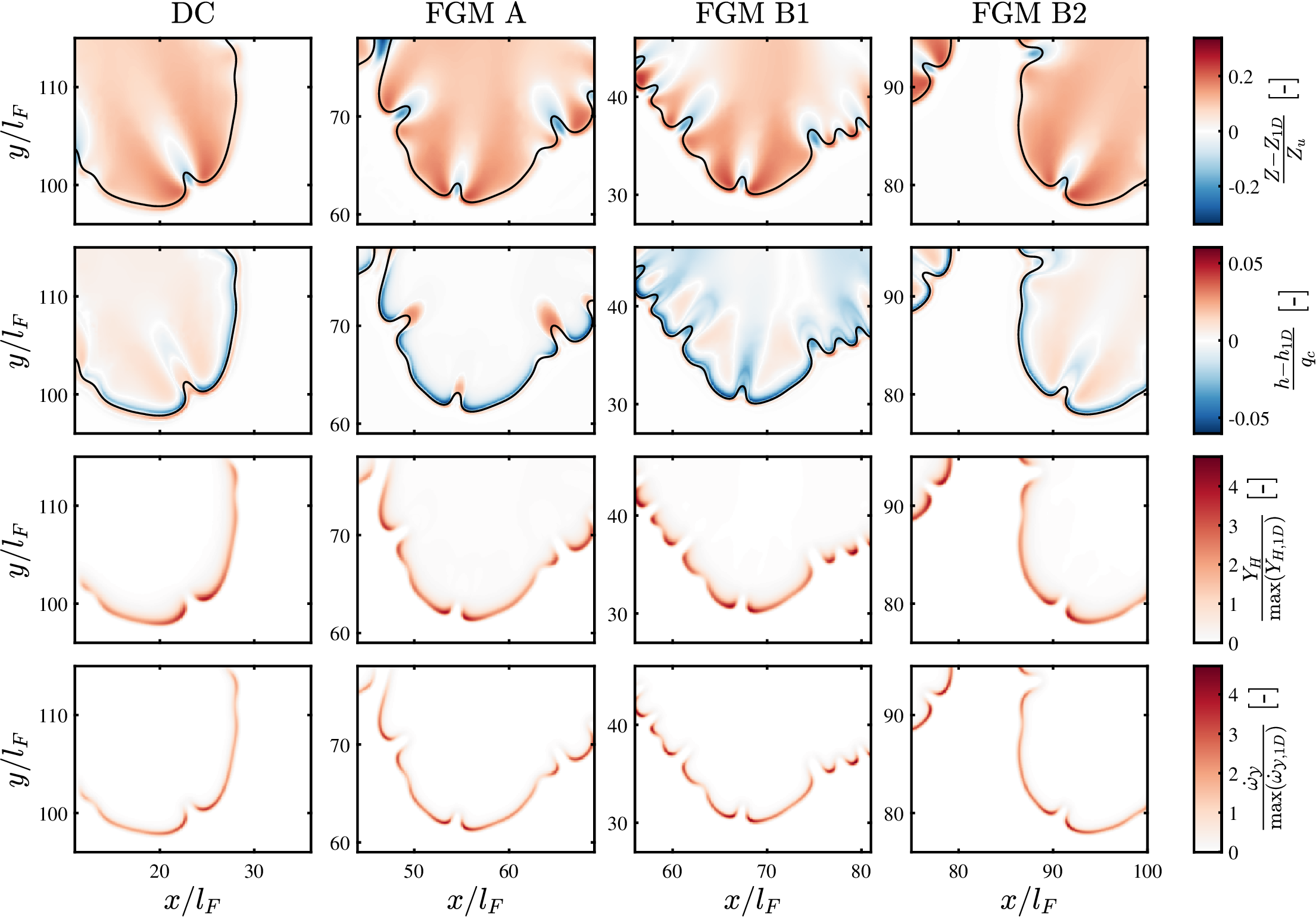}
    \caption{Contour plots of the flame tip / finger regions of the models in consideration. The upper row represents the normalized deviation from the unstretched mixture fraction. The second row represents the normalized deviation from the unstretched enthalpy. The third row represents the normalized H-radical mass fraction and the final row represents the normalized progress variable source term.} 
    \label{fig:contours}
\end{figure}
To further analyze the differences between the models, contour plots of several variables are given in Figure \ref{fig:contours}. The plots depict a zoomed section of a flame tip structure for the considered models: DC, FGM A, FGM B1, and FGM B2. The first row of images shows the normalized deviation from the unstretched mixture fraction for all models considered. The unstretched mixture fraction, $Z_\text{1D}(\mathcal{Y})$, is obtained from the one-dimensional flamelet, and is subtracted from the mixture fraction field to isolate the effect of stretch on the mixture fraction. The resulting value is then normalized with the unburnt mixture fraction value. The effect of preferential diffusion can be observed here: Regions with positive curvature show richer mixtures and vice versa. This results in a lean flame cusp enclosed by two rich flame bulges for the considered models, all of which appear rather similar. The flame tip in FGM B1, however, appears slightly smaller and is enclosed by finer-scale cellular structures, which may be attributed to the shorter unstable wavelengths predicted by the linear stability analysis shown in Figure \ref{fig:LSA}.

The second row of images depicts the normalized deviation from the unstretched enthalpy. Similar to the upper row, the one-dimensional unstretched enthalpy, $h_\text{1D}(\mathcal{Y})$, is subtracted from the enthalpy field, and the resulting value is normalized by the heat of combustion, $q_c$, of the mixture. In the flame bulge, DC shows an increase in enthalpy right before the flame front, followed by a decrease right after the flame front, and then another increase in the post-flame region. In the cusp, there is an increase, followed by a slight decrease in the cusp trail.

In FGM A, enthalpy is not a transported control variable but a dependent variable obtained from the lookup table. The increase right before the flame front and in the cusp is reproduced, and a decrease just after the flame front in the bulge is noticeable. However, in the post-flame region, the enthalpy is zero because it is constant as a function of the mixture fraction. This results in zero enthalpy both before and after the flame, despite changes in the mixture fraction on the burnt side.

In FGM B1, the enthalpy in the flame front as well as behind the flame front shows significantly lower values compared to the DC model. Furthermore, in areas with negative curvature, such as the cusps, there is a decrease in enthalpy instead of an increase, as shown by the DC model. This may suggest that not including the H-radical implicitly in the preferential diffusion of enthalpy limits transport along the flame front, leading to lower enthalpy values in the cusps compared to the DC model.

The enthalpy distribution in FGM B2 is closest to that of DC. Across the entire flame front, there is a slight increase before the flame and a decrease immediately after. In the cusp, the enthalpy is higher compared to FGM B1, possibly because including the H-radical implicitly in the preferential diffusion of enthalpy allows H to diffuse along the flame front, carrying enthalpy to the cusp regions. A slight decrease is also seen in the cusp trail, similar to DC. 

The third row of images shows the H-radical mass fraction, normalized by the maximum value in the unstretched one-dimensional flamelet. In regions of high positive curvature, higher H-radical mass fractions are observed, which is expected due to the richer mixtures in these regions. There seems to be significant variation of H along the flame front, leading to a diffusion flux along it. Since the H-radical carries a significant amount of enthalpy, this could have an impact on the enthalpy diffusion flux along the flame front.

When looking at the source term, normalized by the one-dimensional unstretched maximum value and depicted in the last row, a strong correlation with the H-radical can be observed. By not including the H-radical implicitly in the preferential diffusion of enthalpy, the H-radical cannot diffuse away from the tip, leading to higher H concentrations and a higher source term. These peaks in the reaction rate at the flame tips suggest that these regions play a dominant role in determining the dynamic behavior.

\begin{figure}
    \centering
    \includegraphics[width=\textwidth]{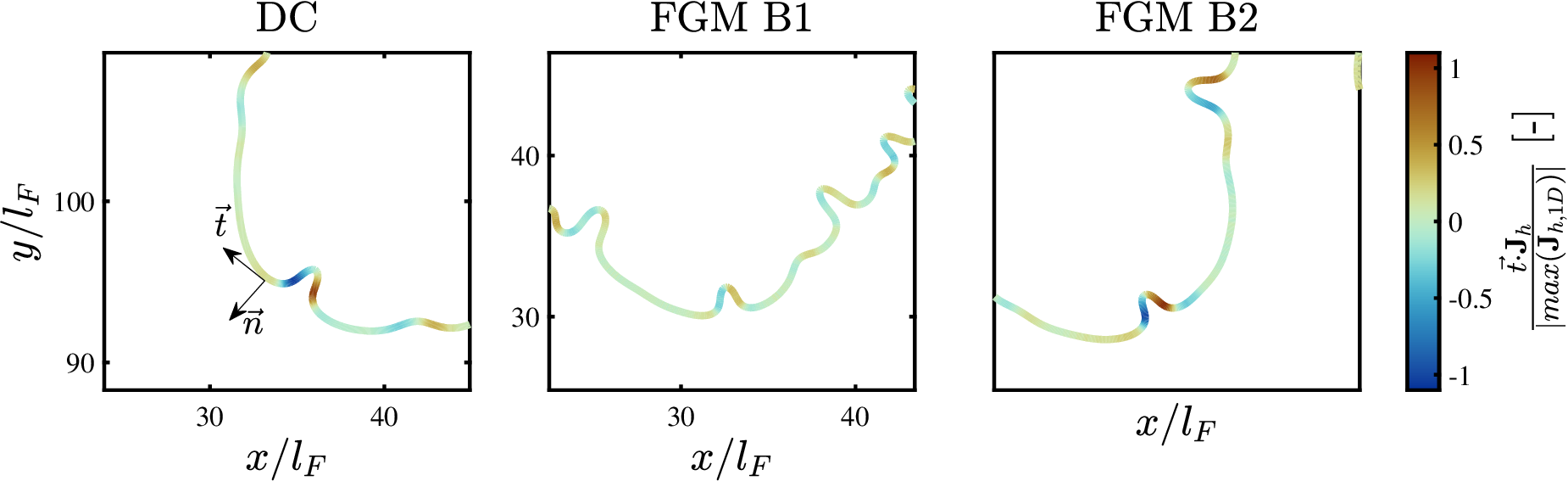}
    \caption{The tangential preferential diffusion flux of enthalpy normalized by its unstretched absolute maximum value conditioned on the progress variable contour line where $\dot{\omega}_\mathcal{Y}$ has its peak value for DC, FGM B1 and FGM B2.}
    \label{fig:tangentialflux}
\end{figure}

Recalling the hypothesis stated in \ref{sec:manifolds}, and according to the observations described above, FGM B1 is not expected to accurately capture the cross-diffusion of enthalpy. The reason for this might be that, in FGM B1, there is no diffusion of the H-radical along the flame front, even though, as shown in Figure \ref{fig:prefdiffcontributionsh}, the H-radical carries a significant amount of enthalpy. To quantify this, the tangential diffusion flux of enthalpy is compared for DC, FGM B1 and FGM B2 in Figure \ref{fig:tangentialflux}. The flame normal vector is defined as
\begin{equation}
    \vec{n} = -\frac{\nabla\mathcal{Y}}{|\nabla\mathcal{Y}|}\;,
\end{equation}
and the tangential vector $\vec{t}$ is perpendicular to $\vec{n}$. For illustration purposes, both vectors are added to the DC plot at a single location on the flame front. The tangential preferential diffusion flux is obtained by taking the dot product between the flame tangential vector and the preferential diffusion flux conditioned on the progress variable isocontour where $\dot{\omega}_\mathcal{Y}$ has its maximum value. Subsequently, this value is normalized by the absolute maximum value of the one-dimensional unstretched preferential diffusion flux of enthalpy. The isocontour is colored with the conditioned tangential preferential diffusion flux of enthalpy and is visualized in Figure \ref{fig:tangentialflux}. One can see that for all three models there is an enthalpy flux towards regions with negative curvature, i.e., the flame cusps. However, FGM B1 seems to significantly under-predict this effect. Looking at the third row of images in Figure \ref{fig:contours}, the gradient of the H-radical is directed towards these flame cusp regions. Since the H-radical has a high specific enthalpy, excluding its effect from the preferential diffusion of enthalpy results in an under-predicted flux toward the flame cusp regions. This is also evident when looking at the enthalpy in the second row of images in Figure \ref{fig:contours}, showing a lower enthalpy value in the flame cusp region for FGM B1 when compared to DC and FGM B2.

\section{Conclusions}
\label{sec:conclusions}

In this work, a planar lean premixed hydrogen flame which exhibits thermo-diffusive instabilities is simulated using the Flamelet-Generated Manifold (FGM) approach. Two manifolds are considered: manifold A, a two-dimensional manifold with progress variable and mixture fraction as control variables, and manifold B, a three-dimensional manifold with enthalpy included as an additional control variable. Choosing enthalpy as a control variable enables the inclusion of heat loss effects on the flame, which is essential for most engineering applications. From an \textit{a-priori} analysis comparing manifold A and B it can be concluded that both manifolds perform well, although manifold B shows slightly improved accuracy over A in terms of normalized error. The newly proposed preferential diffusion model considers only the diffusion fluxes of a limited set of major species.

A one-dimensional preferential diffusion analysis and a linear stability analysis (LSA) suggest that for manifold A, the species H$_2$ and H$_2$O are sufficient, but for manifold B, the H-radical should also be included in the major species set. This is confirmed by a longer simulation in the non-linear regime. Not including the diffusion flux of the H-radical explicitly in the preferential diffusion of enthalpy leads to a poor prediction of the flame's and dynamic behavior. The flame front advances upstream much faster, which is evident from an overprediction in the burning rate and generalized flame surface density. Furthermore, there seem to be smaller cellular structures present, which could be attributed to the small unstable wavelengths that followed from the linear stability analysis. In contrast, including the H-radical diffusion flux in the preferential diffusion of enthalpy results in flame shape, dynamics, and burning rate predictions that closely match the DC benchmark. It is shown that not including the H-radical explicitly leads to a significant underprediction of the tangential enthalpy diffusion flux. The H-radical has a significant tangential component along the flame, and since it carries a significant amount of enthalpy, accurate prediction of enthalpy cross-diffusion requires the explicit inclusion of H-radical diffusion in the enthalpy preferential diffusion term. A final observation is that FGM A also exhibits a slight overprediction in the burning rate, suggesting that enthalpy may play a role as a control variable in accurately capturing thermo-diffusive instabilities. However, further investigation is required to draw definitive conclusions.


To conclude, one can accurately model the preferential diffusion effects that result in a thermo-diffusive unstable flame front with FGM excluding heat losses by only including H$_2$ and H$_2$O in the major species selection, suggesting that cross-diffusion is of less importance here. However, when modelling preferential diffusion with FGM that includes heat loss, cross-diffusion of enthalpy is a critical factor which can be included by incorporating the effect of the H-radical into its preferential diffusion term.

\section{Author contributions}
\textbf{Stijn N.J. Schepers:} Conceptualization, Methodology, Software, Formal analysis, Investigation, Writing - Original Draft, Visualization \textbf{Jeroen A. van Oijen:} Conceptualization, Supervision, Writing - Review \& Editing, Funding acquisition

\section{Declaration of Competing Interest}

The authors declare that they have no known competing financial interests or personal relationships that could have appeared to influence the work reported in this paper. 

\section{Acknowledgements}

This research is part of the HELIOS project "Stable high hydrogen low NOx combustion in full scale gas turbine combustor at high firing temperatures" and is supported by the Clean Hydrogen Partnership and its members.

\section{Declaration of generative AI and AI-assisted technologies in the writing process}

During the preparation of this work the author(s) used ChatGPT by OpenAI in order to check grammar and rewrite some sentences. After using this tool/service, the author(s) reviewed and edited the content as needed and take(s) full responsibility for the content of the publication.



\appendix


 \bibliographystyle{elsarticle-num} 
 \bibliography{references}

\begin{thebibliography}{10}
\expandafter\ifx\csname url\endcsname\relax
  \def\url#1{\texttt{#1}}\fi
\expandafter\ifx\csname urlprefix\endcsname\relax\def\urlprefix{URL }\fi
\expandafter\ifx\csname href\endcsname\relax
  \def\href#1#2{#2} \def\path#1{#1}\fi

\bibitem{Altantzis2011DetailedFlames}
C.~Altantzis, C.~E. Frouzakis, A.~G. Tomboulides, S.~G. Kerkemeier, K.~Boulouchos, {Detailed numerical simulations of intrinsically unstable two-dimensional planar lean premixed hydrogen/air flames}, Proceedings of the Combustion Institute 33~(1) (2011) 1261--1268.

\bibitem{Altantzis2012HydrodynamicFlames}
C.~Altantzis, C.~E. Frouzakis, A.~G. Tomboulides, M.~Matalon, K.~Boulouchos, {Hydrodynamic and thermodiffusive instability effects on the evolution of laminar planar lean premixed hydrogen flames}, Journal of Fluid Mechanics 700 (2012) 329--361.

\bibitem{Kadowaki2005TheInstability}
S.~Kadowaki, H.~Suzuki, H.~Kobayashi, {The unstable behavior of cellular premixed flames induced by intrinsic instability}, Proceedings of the Combustion Institute 30~(1) (2005) 169--176.

\bibitem{Berger2019CharacteristicFlames}
L.~Berger, K.~Kleinheinz, A.~Attili, H.~Pitsch, {Characteristic patterns of thermodiffusively unstable premixed lean hydrogen flames}, Proceedings of the Combustion Institute 37~(2) (2019) 1879--1886.

\bibitem{Berger2023FlameFlames}
L.~Berger, M.~Grinberg, B.~J{\"{u}}rgens, P.~E. Lapenna, F.~Creta, A.~Attili, H.~Pitsch, {Flame fingers and interactions of hydrodynamic and thermodiffusive instabilities in laminar lean hydrogen flames}, Proceedings of the Combustion Institute 39~(2) (2023) 1525--1534.

\bibitem{Creta2020PropagationInstabilities}
F.~Creta, P.~E. Lapenna, R.~Lamioni, N.~Fogla, M.~Matalon, {Propagation of premixed flames in the presence of Darrieus–Landau and thermal diffusive instabilities}, Combustion and Flame 216 (2020) 256--270.

\bibitem{Kadowaki2005NumericalInteraction}
S.~Kadowaki, T.~Hasegawa, {Numerical simulation of dynamics of premixed flames: Flame instability and vortex-flame interaction}, Progress in Energy and Combustion Science 31~(3) (2005) 193--241.

\bibitem{Wen2024ThermodiffusivelyPatterns}
X.~Wen, L.~Berger, L.~Cai, A.~Parente, H.~Pitsch, {Thermodiffusively unstable laminar hydrogen flame in a sufficiently large 3D computational domain – Part I: Characteristic patterns}, Combustion and Flame 263 (5 2024).

\bibitem{Matalon1982FlamesDiscontinuities}
M.~Matalon, B.~J. Matkowsky, {Flames as gasdynamic discontinuities}, Journal of Fluid Mechanics 124 (1982) 239--259.

\bibitem{Matalon2003HydrodynamicOrders}
M.~Matalon, C.~Cui, J.~K. Bechtold, {Hydrodynamic theory of premixed flames: Effects of stoichiometry, variable transport coefficients and arbitrary reaction orders}, Journal of Fluid Mechanics 487 (2003) 179--210.

\bibitem{Howarth2022AnFlames}
T.~L. Howarth, A.~J. Aspden, {An empirical characteristic scaling model for freely-propagating lean premixed hydrogen flames}, Combustion and Flame 237 (3 2022).

\bibitem{Howarth2023Thermodiffusively-unstablePoints}
T.~L. Howarth, E.~F. Hunt, A.~J. Aspden, {Thermodiffusively-unstable lean premixed hydrogen flames: Phenomenology, empirical modelling, and thermal leading points}, Combustion and Flame 253 (7 2023).

\bibitem{Aspden2016}
A.~J. Aspden, M.~S. Day, J.~B. Bell, Three-dimensional direct numerical simulation of turbulent lean premixed methane combustion with detailed kinetics, Combustion and Flame 166 (2016) 266--283.

\bibitem{Aspden2017}
A.~J. Aspden, A numerical study of diffusive effects in turbulent lean premixed hydrogen flames, Proceedings of the Combustion Institute 36 (2017) 1997--2004.
\newblock \href {https://doi.org/10.1016/j.proci.2016.07.053} {\path{doi:10.1016/j.proci.2016.07.053}}.

\bibitem{vanOijen2000ModellingManifolds}
J.~A. van Oijen, L.~P. de~Goey, {Modelling of premixed laminar flames using flamelet-generated manifolds}, Combustion Science and Technology 161~(1) (2000) 113--137.

\bibitem{VanOijen2016}
J.~A. van Oijen, A.~Donini, R.~J. Bastiaans, J.~H. ten Thije~Boonkkamp, L.~P. de~Goey, {State-of-the-art in premixed combustion modeling using flamelet generated manifolds}, Progress in Energy and Combustion Science 57 (2016) 30--74.

\bibitem{Pierce2004Progress-variableCombustion}
C.~D. Pierce, P.~Moin, {Progress-variable approach for large-eddy simulation of non-premixed turbulent combustion}, Journal of Fluid Mechanics 504~(March 2002) (2004) 73--97.

\bibitem{Gicquel2000LAMINARDIFFUSION}
O.~Gicquel, N.~Darabiha, D.~Th{\'{e}}venin, {Laminar Premixed Hydrogen/Air Counterflow Flame Simulations Using Flame Prolongation of ILDM with Differential Diffusion}, Proceedings of the Combustion Institute 28 (2000) 1901--1908.

\bibitem{vanOijen2002ModellingMethod}
J.~A. van Oijen, L.~P.~H. de~Goey, {Modelling of premixed counterflow flames using the flamelet-generated manifold method}, Combustion Theory and Modelling 6~(3) (2002) 463--478.

\bibitem{DeSwart2010InclusionManifolds}
J.~A. De~Swart, R.~J. Bastiaans, J.~A. van Oijen, L.~P.~H. de~Goey, R.~S. Cant, {Inclusion of preferential diffusion in simulations of premixed combustion of hydrogen/methane mixtures with flamelet generated manifolds}, Flow, Turbulence and Combustion 85~(3-4) (2010) 473--511.

\bibitem{Donini2015DifferentialFlames}
A.~Donini, R.~J. Bastiaans, J.~A. van Oijen, L.~P. de~Goey, {Differential diffusion effects inclusion with flamelet generated manifold for the modeling of stratified premixed cooled flames}, Proceedings of the Combustion Institute 35~(1) (2015) 831--837.

\bibitem{Regele2013AFlames}
J.~D. Regele, E.~Knudsen, H.~Pitsch, G.~Blanquart, {A two-equation model for non-unity Lewis number differential diffusion in lean premixed laminar flames}, Combustion and Flame 160~(2) (2013) 240--250.

\bibitem{Schlup2019ReproducingFlames}
J.~Schlup, G.~Blanquart, {Reproducing curvature effects due to differential diffusion in tabulated chemistry for premixed flames}, Proceedings of the Combustion Institute 37~(2) (2019) 2511--2518.

\bibitem{Abtahizadeh2015DevelopmentFlames}
E.~Abtahizadeh, P.~de~Goey, J.~van Oijen, {Development of a novel flamelet-based model to include preferential diffusion effects in autoignition of CH4/H2 flames}, Combustion and Flame 162~(11) (2015) 4358--4369.

\bibitem{Kai2023LESEffect}
R.~Kai, T.~Tokuoka, J.~Nagao, A.~L. Pillai, R.~Kurose, {LES flamelet modeling of hydrogen combustion considering preferential diffusion effect}, International Journal of Hydrogen Energy 48~(29) (2023) 11086--11101.

\bibitem{Zhang2021LargeStretch}
W.~Zhang, S.~Karaca, J.~Wang, Z.~Huang, J.~van Oijen, {Large eddy simulation of the Cambridge/Sandia stratified flame with flamelet-generated manifolds: Effects of non-unity Lewis numbers and stretch}, Combustion and Flame 227 (2021) 106--119.

\bibitem{Zhang2023Large-eddyManifolds}
W.~Zhang, W.~Han, J.~Wang, Z.~Huang, W.~Jin, J.~van Oijen, {Large-eddy simulation of the Darmstadt multi-regime turbulent flame using flamelet-generated manifolds}, Combustion and Flame 257 (2023) 113001.

\bibitem{Mukundakumar2021AFlames}
N.~Mukundakumar, D.~Efimov, N.~Beishuizen, J.~van Oijen, {A new preferential diffusion model applied to FGM simulations of hydrogen flames}, Combustion Theory and Modelling 25~(7) (2021) 1245--1267.

\bibitem{Bottler2022FlameletMixtures}
H.~B{\"{o}}ttler, X.~Chen, S.~Xie, A.~Scholtissek, Z.~Chen, C.~Hasse, {Flamelet modeling of forced ignition and flame propagation in hydrogen-air mixtures}, Combustion and Flame 243 (9 2022).

\bibitem{Bottler2023FlameletFlames}
H.~B{\"{o}}ttler, H.~Lulic, M.~Steinhausen, X.~Wen, C.~Hasse, A.~Scholtissek, {Flamelet modeling of thermo-diffusively unstable hydrogen-air flames}, Proceedings of the Combustion Institute 39~(2) (2023) 1567--1576.

\bibitem{Bilger1990OnFlames}
R.~W. Bilger, S.~H. St{\aa}rner, {On reduced mechanisms for Methane-Air combustion in non-premixed flames}, Combustion and Flame 80 (1990) 135--149.

\bibitem{Fortes2024AnalysisModel}
E.~M. Fortes, E.~J. P{\'{e}}rez-S{\'{a}}nchez, A.~Both, T.~Grenga, D.~Mira, {Analysis of thermodiffusive instabilities in hydrogen premixed flames using a tabulated flamelet model}, Combustion and Flame (11 2024).
\newblock \href {http://arxiv.org/abs/2411.03526} {\path{arXiv:2411.03526}}.

\bibitem{ThierryPoinsot2005TheoreticalCombustion}
T.~Poinsot, D.~Veynante, {Theoretical and Numerical Combustion}, 2nd Edition, Edwards, 2005.

\bibitem{Burke2011ComprehensiveCombustion}
M.~P. Burke, M.~Chaos, Y.~Ju, F.~L. Dryer, S.~J. Klippenstein, {Comprehensive H2/O2 kinetic model for high‐pressure combustion}, International Journal of Chemical Kinetics 44~(7) (2011) 444--474.

\bibitem{70136d09e89b41be89fb05427ca596b8}
L.~Somers, The simulation of flat flames with detailed and reduced chemical models, Ph.D. thesis, Eindhoven University of Technology, Mechanical Engineering (1994).

\bibitem{Matalon2018}
M.~Matalon, The darrieus-landau instability of premixed flames, Fluid Dynamics Research 50 (8 2018).

\end{thebibliography}





\end{document}